\documentclass[11pt]{article}
\usepackage{amsmath}
\usepackage{amssymb}
\usepackage{graphicx}
\usepackage{epsfig}
\usepackage{graphics}
\usepackage{color}

\usepackage{amsmath,amsfonts,latexsym,array,lscape,amssymb,amsthm,cases,empheq}
\usepackage{indentfirst,titlepic}
\usepackage{graphicx,tikz,caption,subcaption}
\usepackage{url}  % command \url from this package allows a long url to break in sensible places
\usepackage{cite} % changes reference [1,2,3] into [1-3]
\numberwithin{equation}{section}

\newcommand{\be}{\begin{equation}}
\newcommand{\ee}{\end{equation}}

\newcommand{\blue}{\color{\blue}}

\begin{document}

\title{\bf Long internal ring waves in a two-layer fluid\\ with an upper-layer current}
%models for nonlinear long internal waves}
\author{Karima Khusnutdinova \\[2ex]
{\small Department of Mathematical Sciences, Loughborough University,}\\
{\small  Loughborough LE11 3TU, UK} \\[1ex]
{\small Email: K.Khusnutdinova@lboro.ac.uk}\\[2ex]
{\it Dedicated to Efim Pelinovsky on the occasion of his 75th Birthday.}}
\date{}
\maketitle

\begin{abstract}
We consider a two-layer fluid with a depth-dependent upper-layer current (e.g. a river inflow, an exchange flow in a strait, or a wind-generated current). In the rigid-lid approximation, we find the necessary singular solution of the nonlinear first-order ordinary differential equation responsible for the adjustment of the speed of the long interfacial ring wave in different directions in terms of the hypergeometric function. This allows us to obtain an analytical description of the wavefronts and vertical structure of the ring waves for a large family of the current profiles and to illustrate their dependence on the density jump and the type and the strength of the current. In the limiting case of a constant upper-layer current  we obtain a 2D ring waves' analogue of the long-wave instability criterion for plane interfacial waves. On physical level, the presence of instability for a sufficiently strong current manifests itself already in the stable regime in the squeezing of the wavefront of the interfacial ring wave in the direction of the current. We show that similar phenomenon can also take place for other, depth-dependent currents in the family.

\end{abstract}

\section{Introduction}

Long-wave models of the Korteweg-de Vries (KdV) type have found numerous useful applications in the studies of the oceanic nonlinear surface and internal waves  (see, for example, \cite{GOSS, HM, GPT, GPTK} and references therein). Two-dimensional generalisations of these models developed in the context of fluids include versions of the Kadomtsev-Petviashvili (KP) equation for water waves in Cartesian \cite{AS}, cylindrical \cite{J1} and elliptic-cylindrical \cite{KKMS}  geometries and internal waves, with a possible background shear flow, in Cartesian geometry (see \cite{G} and references therein), as well as cylindrical Korteweg-de Vries (cKdV)-type models for surface waves without shear flow \cite{M} and on a shear flow \cite{J2}, and internal waves without shear flow \cite{L} and on a shear flow \cite{KZ1, KZ2}. The latter models are relevant to the description of the nearly annular internal waves generated in straits, river-sea interaction zones as well as waves scattered by localised topographic features (see \cite{VSSGL, NM, VSPI, KZ1} and references therein). Some relevant initial-value problems have been considered in \cite{WZ, RRS, MS, KZ2, G}. 

Several recent studies have predicted various effects of shear flows on two-dimensional surface ring and ship waves (see \cite{J2, E1, E2, KZ1, LE} and references therein) and some of these predictions have been recently confirmed in laboratory experiments \cite{SEE}. The study of the effects of a piecewise-constant shear flow on long surface and interfacial ring waves in a two-layer fluid has  shown a striking difference in the shapes of the wavefronts of these waves: while the surface ring waves were elongated in the direction of the current, the interfacial ring waves were squeezed in that direction \cite{KZ1}. Overall, the study of the effects of the shear flow on internal ring waves is in its infancy. It the present paper we aim to build a rather large family of current profiles amenable to theoretical analysis which could be used to approximate some currents present in natural settings, and to elucidate their effects on the long internal ring waves.

\section{Modal equations for ring waves}

In this section we overview the derivation of the far-field set of modal equations for long ring waves in a stratified fluid over a parallel depth-dependent shear flow derived and studied in \cite{KZ1,KZ2}. A ring wave propagates in an inviscid incompressible fluid, described by the set of Euler equations:
\begin{eqnarray}
&& \rho (u_t + u u_x + v u_y + w u_z) + p_x = 0, \label{1} \\
&& \rho (v_t + u v_x + v v_y + w v_z) + p_y = 0, \label{2} \\
&& \rho (w_t + u w_x + v w_y + w w_z) + p_z + \rho g = 0, \label{3} \\
&& \rho_t + u \rho_x + v \rho_y + w \rho_z = 0, \label{4} \\
&& u_x + v_y + w_z = 0, \label{5} 
\end{eqnarray}
subject to the free surface and rigid bottom boundary conditions:
\begin{eqnarray}
&&w = h_t + u h_x + v h_y \quad \mbox{at} \quad z = h(x,y,t), \label{6} \\
&&p = p_a \quad \mbox{at} \quad z = h(x,y,t), \label{7} \\
&&w = 0 \quad \mbox{at} \quad z = 0. \label{8}
\end{eqnarray}
Here, $u,v,w$ are the velocity components in $x,y,z$ directions respectively, $p$ is the pressure, $\rho$ is the density, $g$ is the gravitational acceleration, $z = h(x,y,t)$ is the free surface height ($z = 0$ at the bottom), and $p_a$ is the constant atmospheric pressure at the surface. We assume that in the basic state $u_0 = u_0(z), ~ v_0 = w_0 = 0, ~p_{0z} = - \rho_0 g, ~h = h_0$. Here $u_0(z)$ is a horizontal  shear flow in the $x$-direction, and  $\rho_0 = \rho_0(z)$ is a stable background density stratification. The vertical particle displacement $\zeta$ is used as an additional dependent variable, which is defined by the equation
\begin{equation}
\zeta_t + u \zeta_x + v \zeta_y + w \zeta_z = w, \label{9}
\end{equation}
subject to the surface boundary condition 
\begin{equation}
\zeta = h - h_0 \quad  \mbox{at} \quad z = h(x,y,t), \label{10}
\end{equation}
where $h_0$ is the unperturbed fluid depth.

%We aim to derive an amplitude equation for the amplitudes of the long surface and internal waves. %Thus, 
The problem is considered using the following non-dimensional  set of variables:
\begin{eqnarray*}
&&x \to \lambda x, \quad y \to \lambda y, \quad z \to h_0 z, \quad t \to \frac{\lambda}{c^*}t, \\
&& u \to c^* u, \quad v \to c^* v, \quad w \to \frac{h_0 c^*}{\lambda} w, \\
&&(\rho_0, \rho) \to \rho^*( \rho_0, \rho),  \quad h \to h_0 + a \eta, \\
&& p \to p_a + \int_{z}^{h_0} \rho^* \rho_0(s) g ~\mathrm{d} s + \rho^* g h_0 p,
\end{eqnarray*}
where $\lambda$ is the wave length, $a$ is the wave amplitude, $c^*= \sqrt{g h_0}$ is the  long-wave speed of surface waves, 
%($\sqrt{g h_0}$ or $h^* N^*$, respectively, where $N^*$ is a typical value of the buoyancy frequency, %and $h^*$ is a typical depth of the stratified layer), 
$\rho^*$ is the dimensional reference density of the fluid, while $\rho_0(z)$ is the non-dimensional function describing stratification in the basic state, and $\eta = \eta(x,y,t)$ is the non-dimensional free surface perturbation. 
Non-dimensionalisation leads to the appearance of two small parameters in the problem, the amplitude parameter $\varepsilon = a/h_0$ and the wavelength parameter $\delta = h_0/\lambda$.   The maximal balance condition $\delta^2 = \varepsilon$ has been imposed in \cite{KZ1}.
% although this is not the necessary condition. Indeed, 
%Variables can be scaled further to replace $\delta^2$ with $\varepsilon$ in the equations  \cite{J_book}. 

% Thus, it is natural to non-dimensionalise the general problem formulation, including both surface and internal waves,  using the parameters of the faster surface waves, and measure the speeds of the internal waves as fractions of the surface wave speed, etc. However, if one is primarily interested in the study of internal waves, it is more natural to use the typical speed of the internal waves. 

The problem is then solved in the moving cylindrical coordinate frame (moving at a constant speed $c$: a natural choice is the speed of the shear flow at the bottom, as follows from the derivation). We consider deviations from the basic state and use the same notations $u$ and $v$ for the projections 
%of the deviations of the speed 
on the new coordinate axis, scaling the appropriate variables by the amplitude parameter $\varepsilon$:
 \begin{eqnarray*}
&x \to ct + r \cos \theta, ~~ y \to r \sin \theta, ~~ z \to z, ~~ t \to t, \\
&u \to u_0(z) + \varepsilon (u \cos \theta - v \sin \theta), \\
& v \to \varepsilon (u \sin \theta + v \cos \theta), \\
 &w \to \varepsilon w, ~~ p \to  \varepsilon p, ~~ \rho \to \rho_0 + \varepsilon \rho.
 \end{eqnarray*}
% we can arrive the non-dimensional problem formulation.
 
 The modal equations are obtained by looking for a solution of the problem in the form of asymptotic multiple-scales expansions of the form
 $
 \zeta = \zeta_1 + \varepsilon \zeta_2 + \dots,
 $
 and similar expansions for other variables, where
 \begin{equation}
 \zeta_1 = A(\xi, R, \theta) \phi(z, \theta), \label{11}
 \end{equation}
 with the appropriate set of fast and slow variables:
 \begin{eqnarray}
 \xi = r k(\theta) - s t, \quad R = \varepsilon r k(\theta), \quad \theta = \theta,
 \end{eqnarray}
 where  we define $s$ to be the wave speed in the absence of a  shear flow (with $k(\theta) = 1$). When a shear flow is present the function $k(\theta)$ 
 %describes the distortion of the wavefront 
 is responsible for the adjustment of the wave speed in a particular direction, and is to be determined.  
 %In this description, when a shear flow is present, the wave speed in the direction $\theta$ is not equal to $s$, but to $s/k(\theta)$.  
% The choice of the fast and slow variables is similar to that in the derivation of the cKdV-type %equation for the surface waves \citep{Johnson90}, with 
 The formal range of asymptotic validity of the model is defined by the conditions $\xi\sim R\sim {\cal O}(1)$.  To leading order, the wavefront at any fixed moment of time $t$ is described by the equation
%$$rk(\theta)=\text{constant},$$
%\begin{equation}
$rk(\theta)=\mbox{constant},$
%\end{equation}
 and for the sake of definiteness we consider  outward propagating ring waves, requiring that 
% the function 
 $k = k(\theta) > 0$.
 % is strictly positive.  
 
 To leading order, assuming that perturbations of the basic state are caused only by the propagating wave, the motion is described by the solution \cite{KZ1}
 \begin{eqnarray}
&& u_1 = - A \phi u_{0z} \cos \theta - \frac{k F}{k^2 + k^{'2}} A\phi_z, \label{O1_1} \\
&& v_1 = A \phi u_{0z} \sin \theta -  \frac{k' F}{k^2 + k^{'2}} A\phi_z,  \label{O1_2}\\
&& w_1 =  A_{\xi} F \phi,  \label{O1_3} \\
&& p_1 = \frac{\rho_0}{k^2 + k^{'2}} A F^2 \phi_z,  \label{O1_4}  \\
&& \rho_1 = - \rho_{0z} A \phi,  \label{O1_5}  \\
&&\eta_1 = A \phi \quad \mbox{at} \quad z = 1,  \label{O1_6} 
\end{eqnarray}
where the function $\phi = \phi(z, \theta)$ satisfies the following set of  modal equations:
\begin{eqnarray}
&\left (\frac{\rho_0 F^2}{k^2 + k^{'2}} \phi_z\right )_z -  \rho_{0z} \phi = 0, \label{m1} \\
&\frac{F^2}{k^2 + k^{'2}} \phi_z -  \phi = 0 \quad \mbox{at} \quad z=1, \label{m2} \\
&\phi = 0 \quad \mbox{at} \quad z=0, \label{m3}\\
 \mbox{and }~&  F = -s + (u_0 - c) (k \cos \theta - k' \sin \theta). \nonumber
\end{eqnarray}
The speed of the moving coordinate frame $c$ is fixed to be equal to the speed of the shear flow at the bottom, $c =u_0(0)$. Then, $F = -s \ne 0$ at $z = 0$, and the condition $F \phi = 0$ at $z=0$ which appears as a result of the derivation implies (\ref{m3}). 
 %Of course, the physics does not depend on the choice of $c$ (see a discussion in \cite{Johnson90}), but our derivation shows that the mathematical formulation simplifies if we choose $c = u_0(0)$. 
% The values of the wave speed $s$ in the absence of the shear flow, and the pair of functions $\phi(z, \theta)$ and $k(\theta)$,  for a given shear flow,  %constitute solution of the modal equations (\ref{m1}) - (\ref{m3}).  
 %Unlike the surface wave problem considered by \cite{Johnson90}, the exact form of equations for the wave speed $s$, in the absence of a shear flow, and the function $k(\theta)$, for a given shear flow, depend on stratification. 

The amplitude function $A(\xi, R, \theta)$ is then found by considering the equations at ${\cal O}(\varepsilon)$. It satisfies a cylindrical Korteweg - de Vries (cKdV)-type equation \cite{KZ1}
$$
\mu_1 A_R + \mu_2 A A_{\xi} + \mu_3 A_{\xi \xi \xi} + \mu_4 \frac{A}{R} + \mu_5 \frac{A_{\theta}}{R} = 0,
$$
where the coefficients $\mu_i, i = \overline{1, 5}$ are given in terms of solutions of the modal equations (\ref{m1}) - (\ref{m3}) by the following formulae:
\begin{eqnarray}
&&\mu_1 = 2 s \int_0^1 \rho_0 F \phi_z^2 ~ \mathrm{d}z,  \label{c1}\\
&&\mu_2 = - 3 \int_0^1 \rho_0 F^2 \phi_z^3 ~ \mathrm{d}z,   \label{c2}\\
&&\mu_3 = - (k^2 + k'^2) \int_0^1  \rho_0 F^2 \phi^2 ~ \mathrm{d}z,   \label{c3}\\
&&\mu_4 = - \int_0^1 \left \{  \frac{\rho_0 \phi_z^2 k (k+k'')}{(k^2+k'^2)^2} \left ( (k^2-3k'^2) F^2 -
{4 k' (k^2 + k'^2) (u_0-c)\sin \theta} F  \right .  \right . \nonumber \\
&&\left . \left .  - \sin^2 \theta (u_0-c)^2(k^2 + k'^2)^2 \right ) 
  +  \frac{2 \rho_0 k}{k^2 + k'^2} F \phi_z \phi_{z\theta} (k' F + (k^2 + k'^2) (u_0-c) \sin \theta ) \right \} ~ \mathrm{d}z,\qquad   \label{c4}\\
&& \mu_5 = - \frac{2k}{k^2 + k'^2} \int_0^1 \rho_0 F \phi_z^2 [k' F + (u_0-c) (k^2+k'^2) \sin \theta ] ~ \mathrm{d}z.  \label{c5}
\end{eqnarray}

In this paper we are concerned with the analysis of the modal equations for a two-layer fluid with the upper-layer current  in the rigid-lid approximation, which will allow us to describe, to leading order,  the wavefronts and vertical structure of the long interfacial ring waves. Our primary goal is to analyse the sensitivity of the shape of the wavefront to the variability of the background shear flow in the bulk of the fluid layer, but we will also illustrate the 3D modal functions and discuss the onset of the long-wave instability in the limiting case of a piecewise-constant current. In what follows, the free surface condition (\ref{m2}) is eventually replaced with the rigid-lid approximation
 \begin{equation}
 \phi = 0 \quad \mbox{at} \quad z=1.
 \label{rl}
 \end{equation}

\section{Two-layer fluid with an upper-layer current}

\begin{figure}[h]
\begin{center}
\includegraphics[width=6cm]{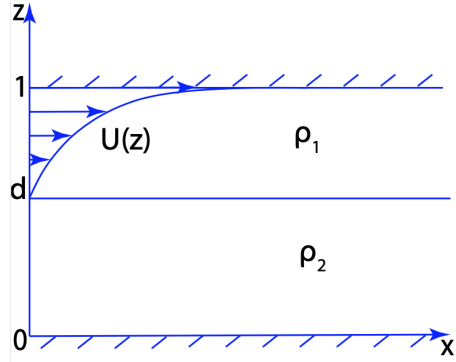} 
\caption{Schematic of the problem formulation.}
\label{fig:schematic}
\end{center}
\end{figure}

We consider a two-layer fluid (see Fig.~\ref{fig:schematic}). The density in the upper layer is $\rho_1$, and the density in the lower payer is $\rho_2$. The flow in the lower layer is assumed to have constant speed, and then without any loss of generality we can assume that this speed is equal to zero. The upper-layer flow is described by the function $U(z)$. Thus,
\begin{eqnarray}
u_0 = \left \{ 
\begin{array}{c}
U(z), \quad \mbox{if} \quad d \le z \le 1, \\
0, \quad \mbox{if} \quad 0 \le z < d.
\end{array}
\right .
\end{eqnarray}
We will assume the continuity of $u_0$ at $z=d$, i.e. $U(d) = 0$. This is a generalisation of the case studied in \cite{J2, J_book} for surface waves in a homogeneous fluid, it is a possible model for a river inflow, an exchange flow in a strait, or a wind-generated current, for example.

The continuous solution of the modal equations (\ref{m1}) - (\ref{m3}) is given (in respective layers) by
\begin{eqnarray}
\phi_1 = \frac{A (k^2 + k'^2)}{\rho_1} \left (1 - \int_z^1 \frac{dz}{F_1^2} \right ), \label{phi1}\\
\phi_2 = \frac{A}{\rho_1 d} \left [1 - (k^2 + k'^2) \int_d^1 \frac{dz}{F_1^2} \right ] z, \label{phi2}
\end{eqnarray}
where $F_1= -s + U(z) (k \cos \theta - k' \sin \theta)$ and $A$ is an arbitrary constant. The derivative of this solution is discontinuous at $z = d$ and must satisfy the jump condition:
\begin{equation}
\frac{F(d)^2}{k^2 + k'^2} [\rho_0 \phi] - [\rho_0] \phi(d) = 0,
\label{jc}
\end{equation}
where $F(d) = -s$, yielding the following nonlinear first-order ordinary differential equation for the function $k(\theta)$:
\begin{equation}
\left (\frac{\rho_1}{\rho_2} - 1 \right ) \left (\int_d^1 \frac{dz}{F_1^2} \right ) (k^2 + k'^2)^2 + \left (1 + \frac{s^2}{d} \int_d^1 \frac{dz}{F_1^2} \right ) (k^2 + k'^2) - \frac{s^2}{d} = 0.
\label{eq_k}
\end{equation}

If there is no current, i.e. $U(z) = 0$, then $F_1 = -s, k = 1, k' = 0$, and the equation (\ref{eq_k}) reduces to an algebraic equation for the speed $s$:
\begin{equation}
s^4 - s^2 - \frac{\rho_1 - \rho_2}{\rho_2} d (1-d) = 0,
\end{equation}
implying
\begin{equation}
s_{1,2}^2 = \frac{1 \pm \sqrt{1 + 4 \frac{\rho_1 - \rho_2}{\rho_2} d (1-d)}}{2},
\label{s}
\end{equation}
giving us the speed of the surface and interfacial ring waves in the absence of any current. 

In the rigid-lid approximation, the solution (\ref{phi1}), (\ref{phi2}) is replaced with
\begin{eqnarray}
\phi_1 = - \frac{A (k^2 + k'^2)}{\rho_1}  \int_z^1 \frac{dz}{F_1^2}, \label{phi1a}\\
\phi_2 = - \frac{A (k^2 + k'^2) z}{\rho_1 d}  \int_d^1 \frac{dz}{F_1^2}, \label{phi2a}
\end{eqnarray}
where $A$ is an arbitrary constant, while the jump condition (\ref{jc}) yields 
\begin{equation}
k^2 + k'^2 = \frac{\rho_1 d + \rho_2 s^2 \int_d^1 \frac{dz}{F_1^2} }{(\rho_2 - \rho_1) d \int_d^1 \frac{dz}{F_1^2} }.
\label{eq_ka}
\end{equation}
When $U(z) = 0$, this equation gives
\begin{equation}
s^2 = \frac{(\rho_2 - \rho_1) d (1-d)}{\rho_1 d + \rho_2 (1-d)},
\label{sa}
\end{equation}
the speed of the interfacial ring wave in the absence of the background current, and in the rigid-lid approximation. The equation (\ref{eq_ka}) and the speed (\ref{sa}) can be formally obtained from (\ref{eq_k}) and (\ref{s}) as an approximation when $\rho_2 - \rho_1 \ll \rho_1, \rho_2$.

\section{Wavefronts and vertical structure}

\begin{figure}[h]
\begin{center}
\includegraphics[width=7cm]{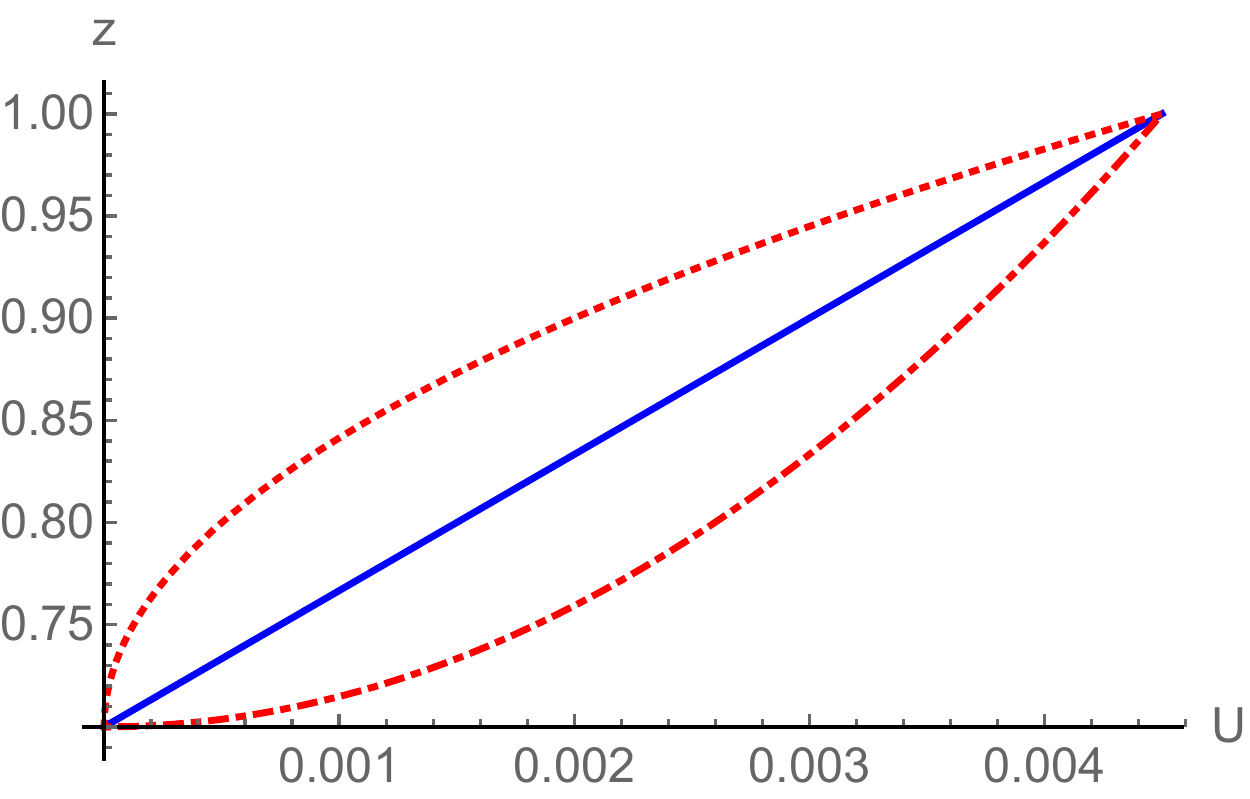} 
\caption{Upper-layer currents $U(z) = \gamma (z-d)$ (blue, solid), $U(z) = \gamma (z-d)^2$ (red, dotted) and $U(z) = \gamma \sqrt{z-d}$ (red, dot-dashed)  for $d = 0.7$ and $U(1) = 0.0045$.}
\label{fig:currents}
\end{center}
\end{figure}

Let us now consider the family of the upper-layer current profiles described by the function
\begin{equation}
U(z) = \gamma (z-d)^{\alpha}, 
\label{U}
\end{equation}
where $\gamma$ and $\alpha$ are some positive constants. In particular, 
the upper-layer current  is shown in Fig.~\ref{fig:currents}  for $\alpha = 1$, $\alpha = \frac 12$ and $\alpha = 2$ for $d= 0.7$ and $\gamma = 0. 015$ when $\alpha = 1$, $\gamma = 0.00821584$ when $\alpha = \frac 12$, $\gamma =  0.05$ when $\alpha = 2$ (all currents have the same strength $U(z) = 0.0045$ on the surface $z=1$).

The equation (\ref{eq_ka}) takes the form
\begin{equation}
k^2 + k'^2 = \frac{\rho_1 d + \rho_2 s^2  I[K (\theta)] }{(\rho_2 - \rho_1) d\  I[K (\theta)]},
\label{eq_kb}
\end{equation}
where 
$$
I[K(\theta)] = \int_d^1 \frac{dz}{F_1^2} = \frac{1-d}{s^2}  {}_2 F_1 \left (2, \frac{1}{\alpha}, 1 + \frac{1}{\alpha}, \frac{\gamma K(\theta)}{s} (1-d)^{\alpha}\right ),
$$
and $K(\theta) = k(\theta) \cos \theta - k' (\theta)  \sin \theta$. Here, the function $I[K(\theta)] $ is given in terms of the hypergeometric function $ {}_2 F_1$.

The general solution of the equation (\ref{eq_kb}) has the form
\begin{equation}
k(\theta) = a \cos \theta + b(a) \sin \theta,
\label{gs}
\end{equation}
where
\begin{equation}
b^2 = - a^2 + \frac{\rho_1 d + \rho_2 s^2\ I(a)}{(\rho_2 - \rho_1) d\ I(a)}
\label{b}
\end{equation}
and
\begin{eqnarray}
I(a) &=& \int_d^1 \frac{dz}{[s - \gamma a (z-d)^{\alpha}]^2} \nonumber \\
      &=& \frac{1-d}{s^2}  {}_2 F_1 \left (2, \frac{1}{\alpha}, 1 + \frac{1}{\alpha}, \frac{\gamma a}{s} (1-d)^{\alpha}\right ).
\end{eqnarray}

The far-field wavefronts of the ring waves at a fixed moment of time are described by the curves
\begin{equation}
 H(r, \theta, t) = r k(\theta) - s t = \mbox{constant},
 \label{H}
\end{equation}
where $k(\theta)$ is the singular solution of the equation (\ref{eq_ka}) (i.e. the envelope of the general solution found by requiring $\frac{d k}{da} = 0$). In what follows all wavefronts are plotted for $r k(\theta) = 50$ ($R \sim 1, \varepsilon \sim 0.02$).

In this general setting we assume that the current is sufficiently weak, so that there exists a part of the wavefront which is able to propagate in the upstream direction, and therefore $\theta \in [-\pi, \pi]$, but we will also discuss other possible regimes for the limiting case of a constant upper-layer current. It is sufficient to define the solution for $\theta \in [0, \pi]$ because of the symmetry of the problem. For the family of currents described by (\ref{U}), the singular solution can be found explicitly  in parametric form
\begin{eqnarray}
&&k(a) = a \cos \theta (a) + b(a) \sin \theta (a),  \label{SS1} \\
&& b = \sqrt{ \frac{\rho_1 d + \rho_2 s^2\ I(a)}{(\rho_2 - \rho_1) d\ I(a)}  - a^2},  \label{SS2} \\
&&\theta (a) = \left \{
\begin{array}{l}
\arctan \frac{2 b (\rho_2 - \rho_1) d\ I^2(a)}{2 a (\rho_2 - \rho_1) d\ I^2(a) + (1-\rho_2 s^2 I(a)) I'(a)} \quad \mbox{if} \quad a \in [a_0, a_{max}]\ (\theta \in [0, \frac{\pi}{2}]), \\
\arctan \frac{2 b (\rho_2 - \rho_1) d\ I^2(a)}{2 a (\rho_2 - \rho_1) d\ I^2(a) + (1-\rho_2 s^2 I(a)) I'(a)} + \pi \quad \mbox{if} \quad a \in [a_{min}, a_0]\  (\theta \in [\frac{\pi}{2}, \pi]);
\end{array}
\right .
\label{SS3}
\end{eqnarray}
where
$$
I'(a) = \frac{1-d}{\alpha a s^2} \left [\frac{1}{[1-\frac{\gamma a}{s} (1-d)^{\alpha}]^2} -  {}_2 F_1 \left (2, \frac{1}{\alpha}, 1 + \frac{1}{\alpha}, \frac{\gamma a}{s} (1-d)^{\alpha}\right )\right ].
$$
Here, the parameter $a$ takes values in the interval $[a_{min}, a_{max}]$ which is found by requiring $b^2 \ge 0$ in (\ref{b}). We consider the outward propagating ring waves, and therefore require $k(\theta) > 0$ for all $\theta$. Then, the interval must contain $a=0$ since $a$ should take both positive and negative values to allow $k(\theta)$ to be positive at both $\theta = 0$ and $\theta = \pi$. The value $a_0 \in [a_{min}, a_{max}]$ is found from the condition 
$$\frac{db(a)}{da} = - \frac{1}{\tan \theta} = 0.$$
 When $\gamma = 0$, $I(a) = \frac{1-d}{s^2} > 0$ yielding $a \in [-1, 1]$ and $a_0 = 0$. By continuity, the real solution will continue to exist at least for a sufficiently small $\gamma$, while the flow might become unstable for some stronger currents \cite{KZ1}.

It is worth noting that in many natural cases the hypergeometric function featured in the solution reduces to elementary functions. In particular, for $\alpha = 1$ (i.e. $U(z) = \gamma (z-d)$), we have
\begin{equation}
 {}_2 F_1 \left (2, 1, 2, \frac{\gamma a}{s} (1-d) \right ) = \frac{1}{1 - \frac{\gamma a}{s} (1-d)},
 \label{HG_lin}
 \end{equation}
 for $\alpha = \frac 12$ (i.e. $U(z) = \gamma \sqrt{z-d}$), 
\begin{equation}
 {}_2 F_1 \left (2, 2, 3, \frac{\gamma a}{s} \sqrt{1-d} \right ) = \frac{2}{1-d}  \left [ \frac{\sqrt{1-d}}{\frac{\gamma a}{s} (1 - \frac{\gamma a }{s}\sqrt{1-d})} + \frac{1}{\gamma^2 a^2} \ln (1 - \frac{\gamma a}{s} \sqrt{1-d}) \right ], 
 \label{HG_root}
 \end{equation}
 and for $\alpha = 2$ (i.e. $U(z) = \gamma (z-d)^2)$,
\begin{equation}
 {}_2 F_1 \left (2, \frac 12, \frac 32, \frac{\gamma a}{s} (1-d)^2 \right ) = \frac{1}{2 (1 - \frac{\gamma a}{s} (1-d)^2)} + \frac{1}{2 \sqrt{\frac{\gamma a}{s}} (1-d)} \arctan [\sqrt {\frac{\gamma a}{s}} (1-d)].
 \label{HG_square}
 \end{equation}
 In the first case, for  $U(z) = \gamma (z-d)$, the singular solution can be rewritten in the form $k = k(\theta)$ as follows
 \begin{equation}
 k(\theta) = \sqrt{1 + \left[ \frac{\rho_1 \gamma s}{2 (\rho_2 - \rho_1)} \right ]^2} - \frac{\rho_1 \gamma s}{2 (\rho_2 - \rho_1)} \cos \theta, \quad \mbox{where} \quad \theta \in [-\pi, \pi].
 \label{ss}
 \end{equation}
 This solution has been used to test the Mathematica code for the general solution (\ref{SS1}) - (\ref{SS3}). The code was then used to plot other figures of the wavefronts of the ring waves shown below. 
 
Following \cite{J2}, we note that the local wave speed in the normal direction to the wavefront, $\frac{\nabla H}{|\nabla H|}$ is given by 
$$- \frac{H_t}{|\nabla H|} = \frac{s}{\sqrt{k^2 + k'^2}}.$$
 Therefore, to avoid the appearance of critical layers we require that
$$
\frac{s}{\sqrt{k^2 + k'^2}} \ne U(z) \cos (\theta + \alpha),
$$
where $\alpha$ is the angle between the radial direction and the normal to the wavefront, and $\cos \alpha = \frac{k}{\sqrt{k^2 + k'^2}}$. This condition is equivalent to 
$$
F_1 = - s + U(z) (k \cos \theta - k' \sin \theta) \ne 0
$$
for $ d \le z \le 1$ since $F_2 = -s \ne 0$. Following \cite{KZ1}, we can obtain a simple sufficient condition for the absence of critical layers.
% in the form
%\begin{equation}
%\gamma (1-d)^\alpha < s.
%\label{est}
%\end{equation}
Indeed, 
$$
F_{1 \theta} =  -  \gamma (z-d)^\alpha  (k + k'') \sin \theta,
$$
where we assume that $\gamma > 0, \alpha >0, d \le z \le 1$ and $k + k'' > 0$ on the selected singular solution (outward propagating wave). Then, $F_1$ has a maximum at $\theta = 0$, and we require that
$$
F_1 \le F_1|_{\theta = 0} = - s + \gamma (z-d)^\alpha k(0) < 0,
$$
implying
$$
\gamma (z-d)^\alpha < \frac{s}{k(0)},
$$
which we replace with a stronger condition
$$
\gamma (z-d)^\alpha \le \gamma (1-d)^\alpha  < s < \frac{s}{k(0)}. 
$$
Thus, in order to avoid the appearance of critical layers, in what follows we impose this constraints on $\gamma$ for the examples of the shear flow.
%\begin{itemize}
%\item for $U(z) = \gamma (z-d)$ (i.e. $\alpha = 1$) we require that $\gamma < \frac{s}{1-d}$,
%\item for $U(z) = \gamma \sqrt{z-d}$ (i.e. $\alpha = \frac 12$) we require that $\gamma < \frac{s}{\sqrt{1-d}}$.
%\end{itemize}

  \begin{figure}[h]
\begin{center}
\includegraphics[height=5.5cm]{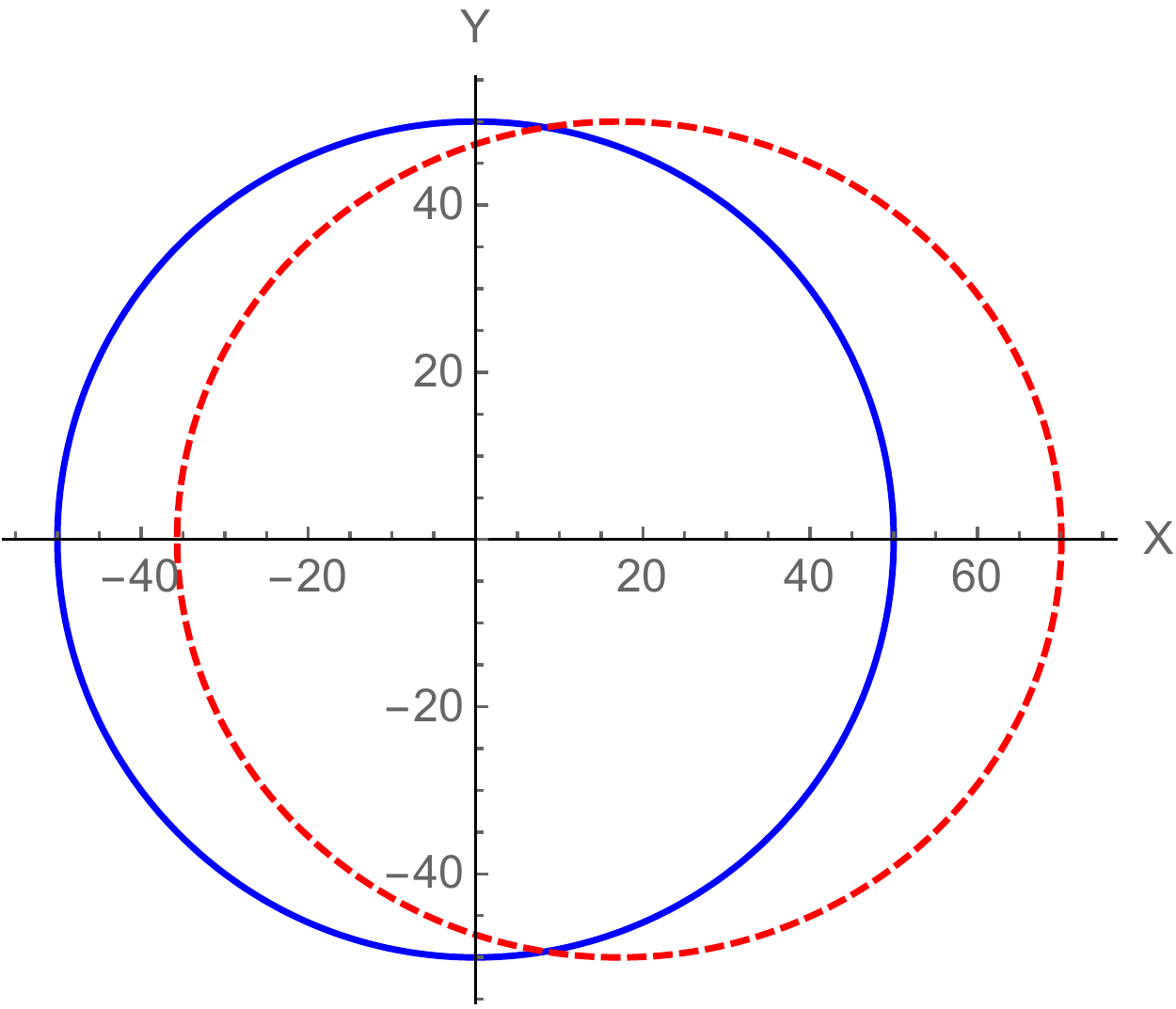} \quad
\includegraphics[height=5.5cm]{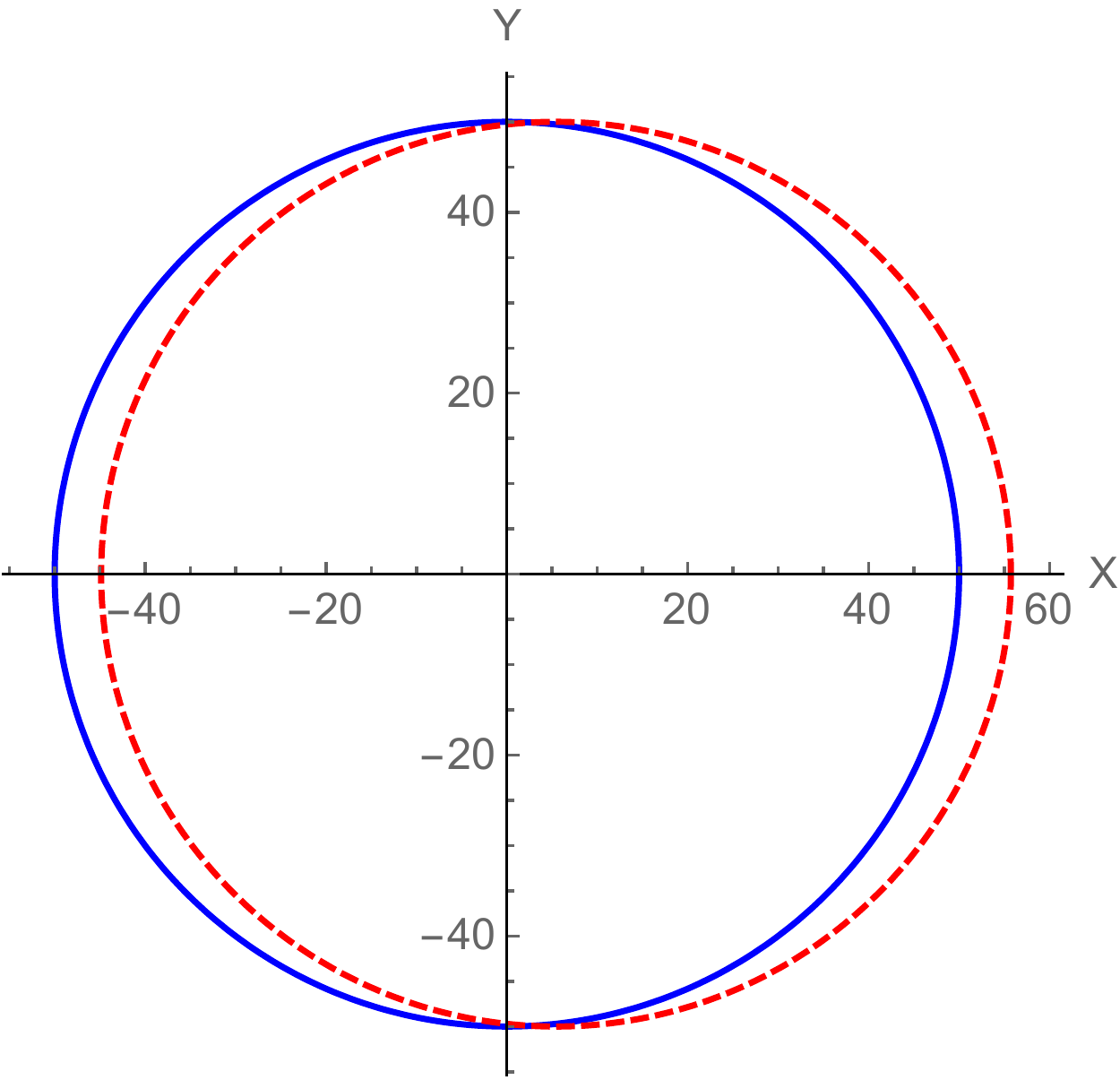}
\caption{Wavefronts of interfacial ring waves for $U(z) = \gamma (z-d)$;   $\rho_1 = 1, \rho_2 = 1.0001$ (left) and  $\rho_2 = 1.001$ (right);  $d = 0.7; \gamma = 0$ (blue, solid), $\gamma = 0.015$ (red, dashed).}
\label{fig:linear}
\end{center}
\end{figure}

 In Fig.~\ref{fig:linear} we show the wavefronts of the interfacial ring waves on the linear current
 $$
 U(z) = \gamma (z-d).
 $$
 The plots on the left are obtained for $\rho_1 = 1, \rho_2 = 1.0001$, while the plots on the right are for a greater density jump, $\rho_1 = 1, \rho_2 = 1.001$, with $d = 0.7$ in all plots. The parameter $\gamma$ takes two values $\gamma_1 = 0$ (blue, solid) and $\gamma_2 = 0.015$ (red, dashed). For the same strength of the shear flow, the wavefronts appear to be convected by the flow and slightly elongated in the direction of the flow, which is more noticeable in the plots on the left, i.e. for the smaller density jump (and therefore slower interfacial waves; indeed, $s_1 =  0.00458251$ and $s_2 = 0.0144892$).

 \begin{figure}[h]
\begin{center}
\includegraphics[height=5.5cm]{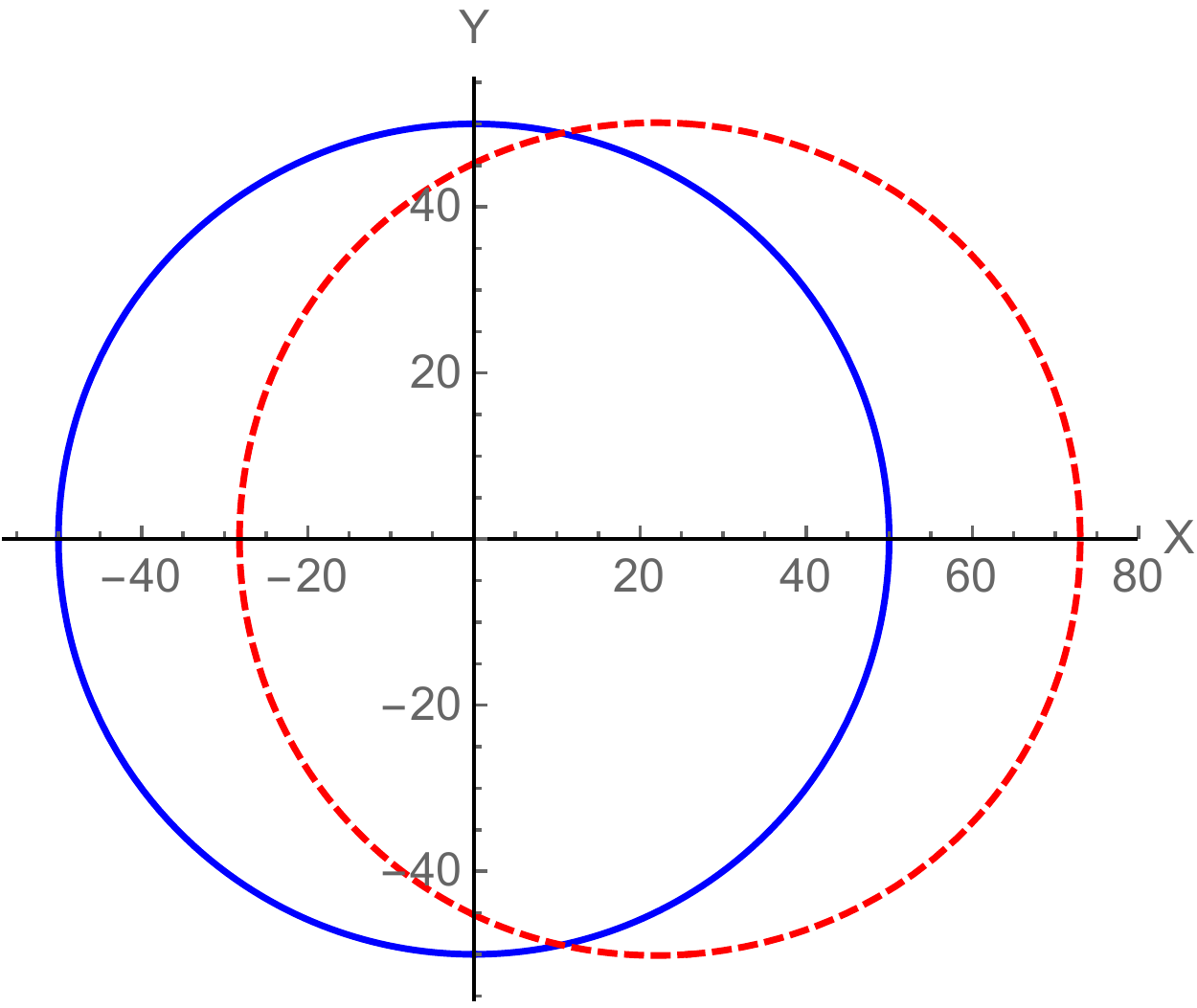} \quad
\includegraphics[height=5.5cm]{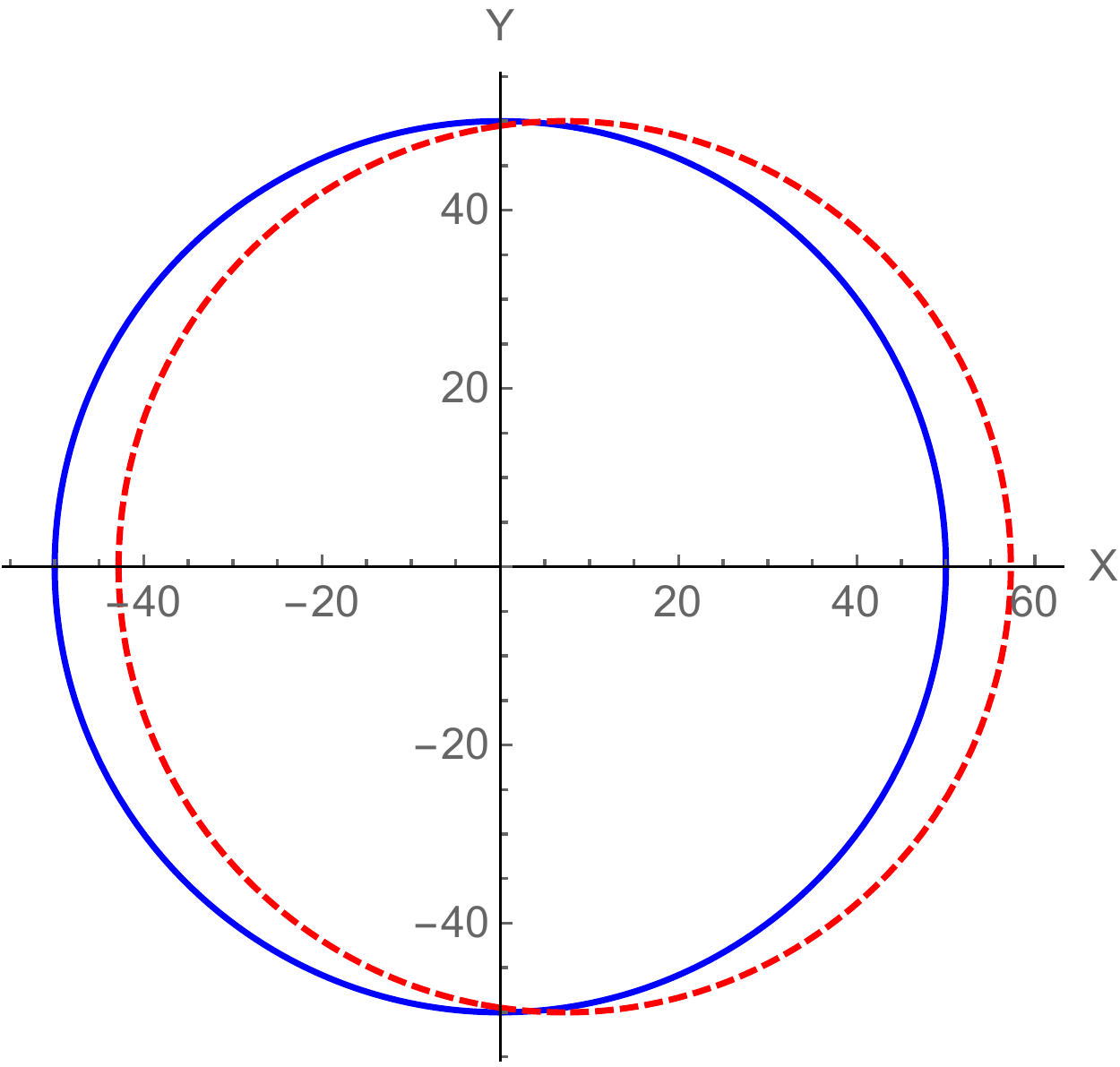}
\caption{Wavefronts of interfacial ring waves for $U(z) = \gamma \sqrt{z-d}$;  $\rho_1 = 1, \rho_2 = 1.0001$ (left) and  $\rho_2 = 1.001$ (right);  $d = 0.7; \gamma = 0$ (blue, solid), $\gamma = 0.00821584$ (red, dashed).}
\label{fig:root}
\end{center}
\end{figure}

  In Fig.~\ref{fig:root} we show the wavefronts of the interfacial ring waves on the  current
 $$
 U(z) = \gamma \sqrt{z-d}.
 $$
 Here again the plots on the left are obtained for $\rho_1 = 1, \rho_2 = 1.0001$, while the plots on the right are for $\rho_1 = 1, \rho_2 = 1.001$, with $d = 0.7$ in all plots. The parameter $\gamma$ takes the values $\gamma_1 = 0$ (blue, solid) and $\gamma_2 = 0.00821584$ (red, dashed). Thus, the first value is the same as before, while the second value is decreased in order to have the same strength of the current $U(z) = 0.0045$ on the surface $z=1$. The wavefronts appear to be mainly convected in the downstream direction which is again much more pronounced  in the plots on the left, i.e. for the smaller density jump.

 \begin{figure}[h]
\begin{center}
\includegraphics[height=5.5cm]{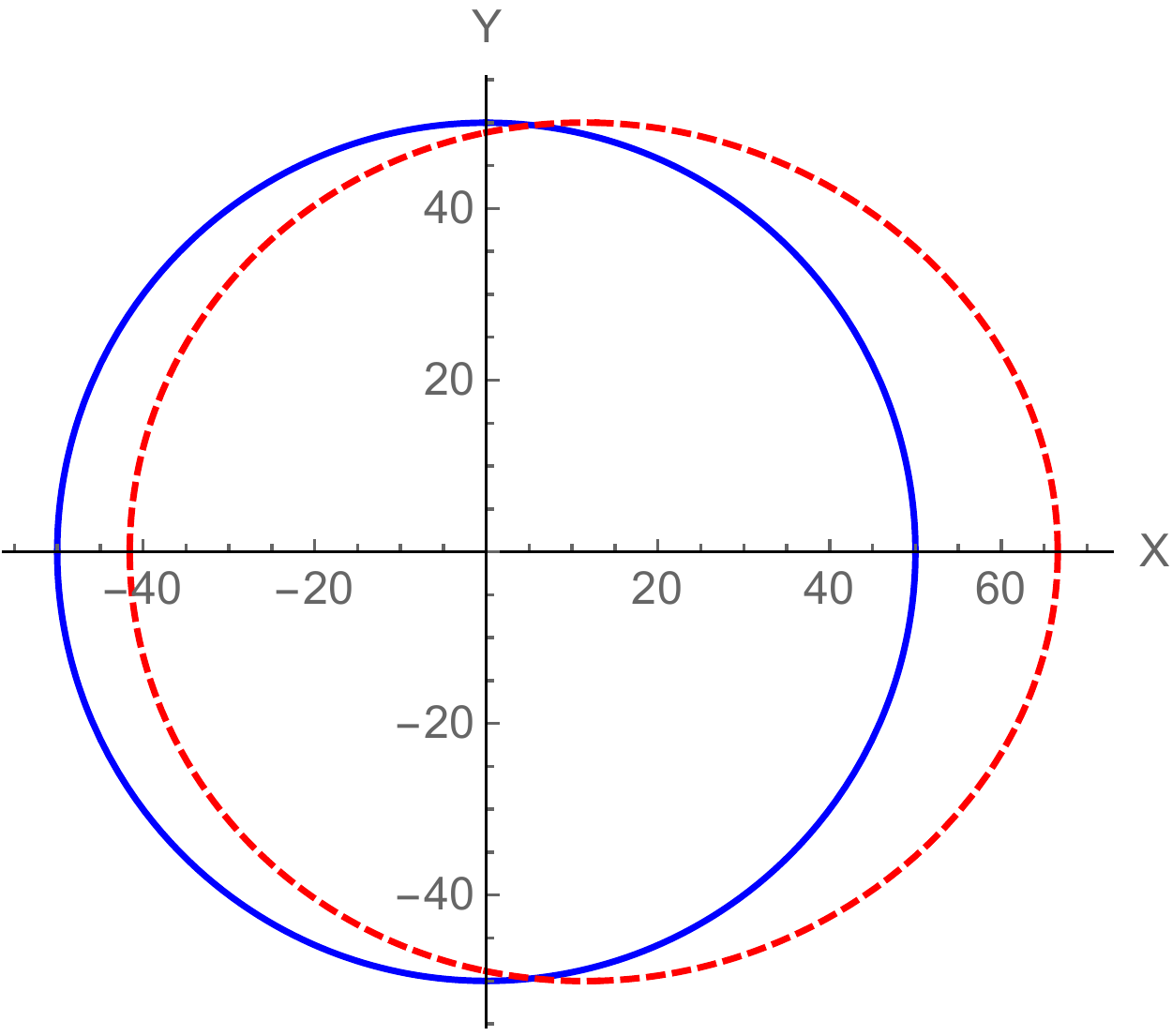} \quad
\includegraphics[height=5.5cm]{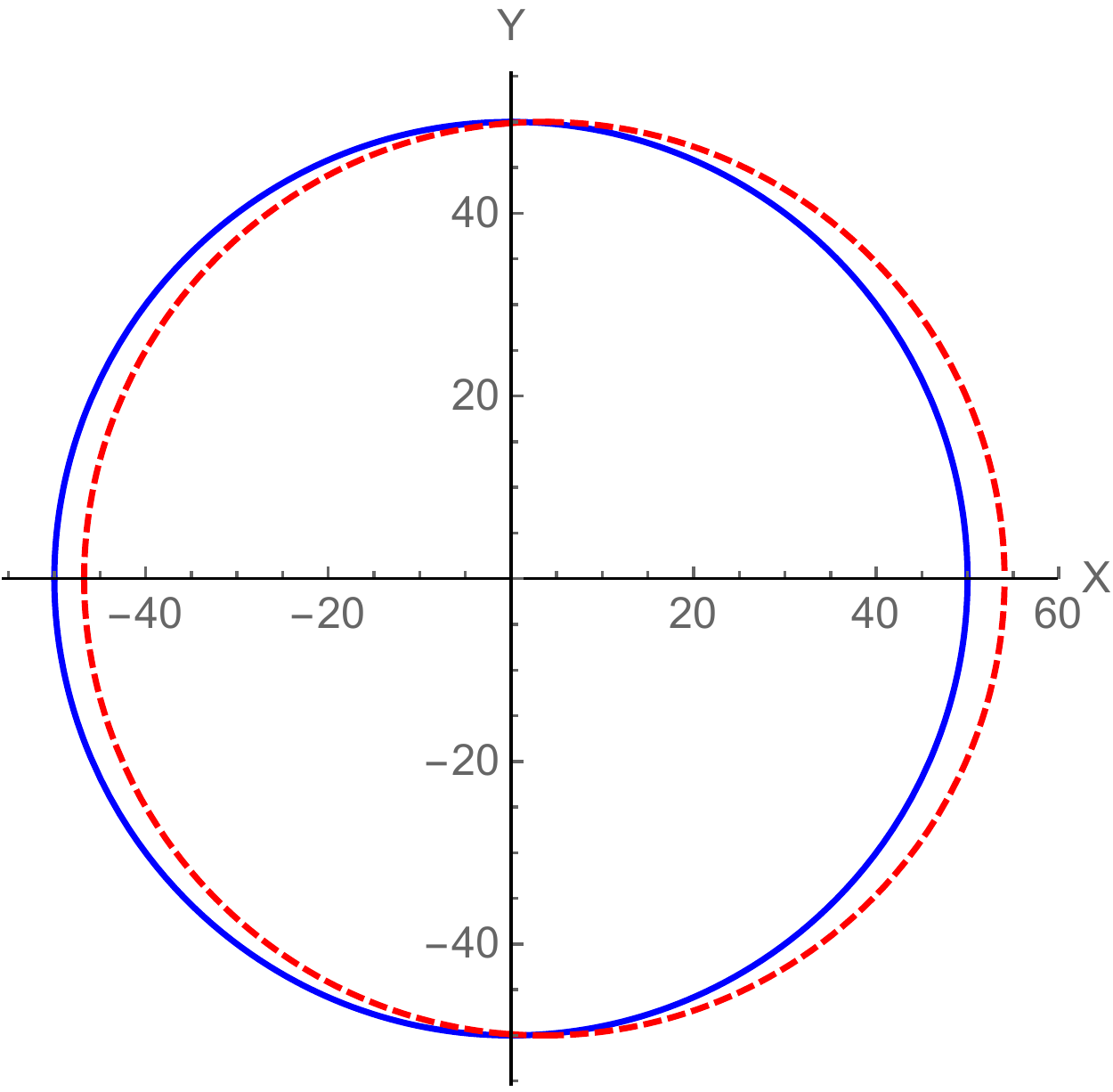}
\caption{Wavefronts of interfacial ring waves for $U(z) = \gamma (z-d)^2$;   $\rho_1 = 1, \rho_2 = 1.0001$ (left) and  $\rho_2 = 1.001$ (right);  $d = 0.7; \gamma = 0$ (blue, solid), $\gamma = 0.05$ (red, dashed).}
\label{fig:square}
\end{center}
\end{figure}

 In Fig.~\ref{fig:square} we show the wavefronts of the interfacial ring waves on the  current
 $$
 U(z) = \gamma (z-d)^2.
 $$
 The plots on the left are obtained for $\rho_1 = 1, \rho_2 = 1.0001$, while the plots on the right are for $\rho_1 = 1, \rho_2 = 1.001$, with $d = 0.7$ in all plots. The parameter $\gamma$ takes the values $\gamma_1 = 0$ (blue, solid) and $\gamma_2 = 0.05$ (red, dashed). Thus, the first value is the same as before, while the second value is increased in order to have the same strength of the current $U(z) = 0.0045$ on the surface $z=1$. The wavefronts appear to be mainly elongated in the downstream direction which is again much more pronounced  in the plots on the left, i.e. for the smaller density jump.

 \begin{figure}[h]
\begin{center}
\includegraphics[height=6.5cm]{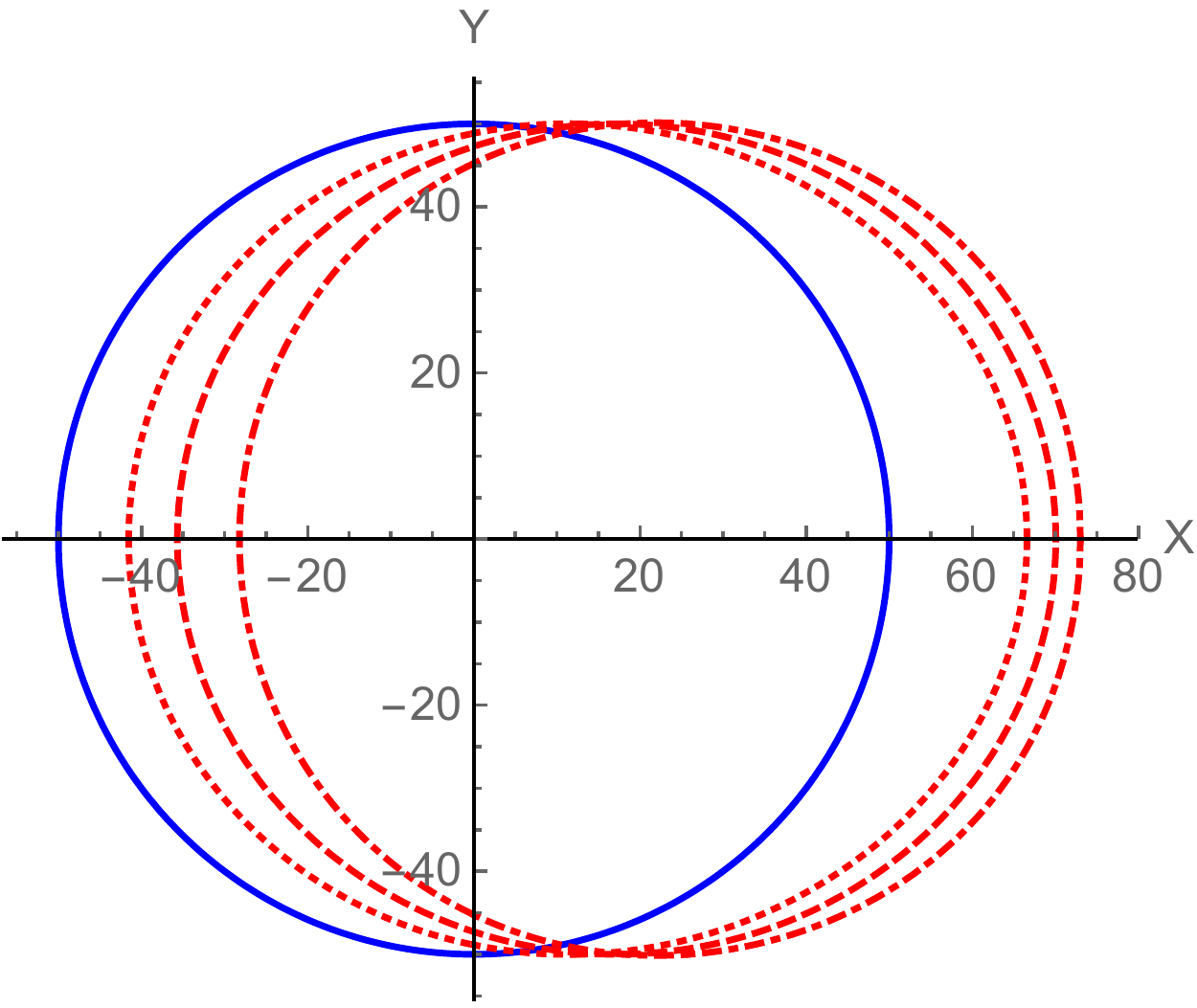}
\caption{Wavefronts of interfacial ring waves for  $\rho_1 = 1, \rho_2 = 1.0001, d = 0.7$; $\gamma = 0$ (blue, solid), $\gamma = 0.05$ for $\alpha = 2$ (red, dotted), $\gamma = 0.015$ for $\alpha = 1$  (red, dashed),  and  $\gamma = 0.00821584$ for $\alpha = \frac 12$  (red, dot-dashed). For all three currents $U(1) = 0.0045$.}
\label{fig:root2}
\end{center}
\end{figure}

In Fig.~\ref{fig:root2} we compare the wavefronts of the interfacial ring waves on the  currents
 $$
 U(z) = \gamma (z-d)^\alpha \quad \mbox{where}  \quad  \alpha = \frac 12, 1, 2
 $$
 for the same set of parameters. Here, $\rho_1 = 1, \rho_2 = 1.0001, d = 0.7$ and  $\gamma = 0$ (blue, solid), $\gamma = 0.05$ for $\alpha = 2$ (red, dotted),  $\gamma = 0.015$ for $\alpha = 1$ (red, dashed)  and $\gamma = 0.00821584$ for $\alpha = \frac 12$ (red, dot-dashed). All currents have the same strength $U(z) = 0.0045$ on the surface $z=1$.
It appears that the current with $\alpha = \frac 12$ convects the ring further downstream than the currents with $\alpha = 1$ and $\alpha = 2$, while the current with $\alpha = 2$ has a stronger effect on the shape of the wavefront elongating it in the direction of the shear flow.

\begin{figure}[b]
\begin{center}
\includegraphics[height=3.4cm]{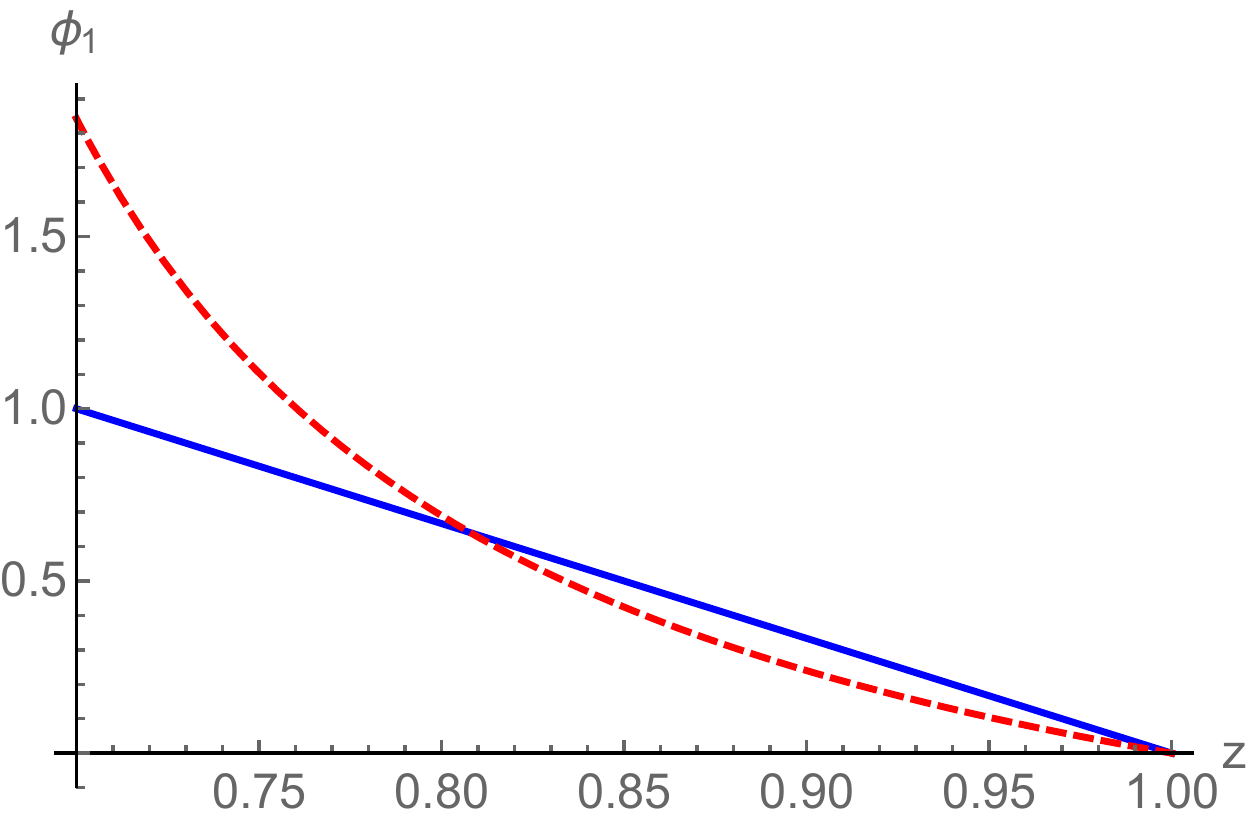}
\includegraphics[height=3.4cm]{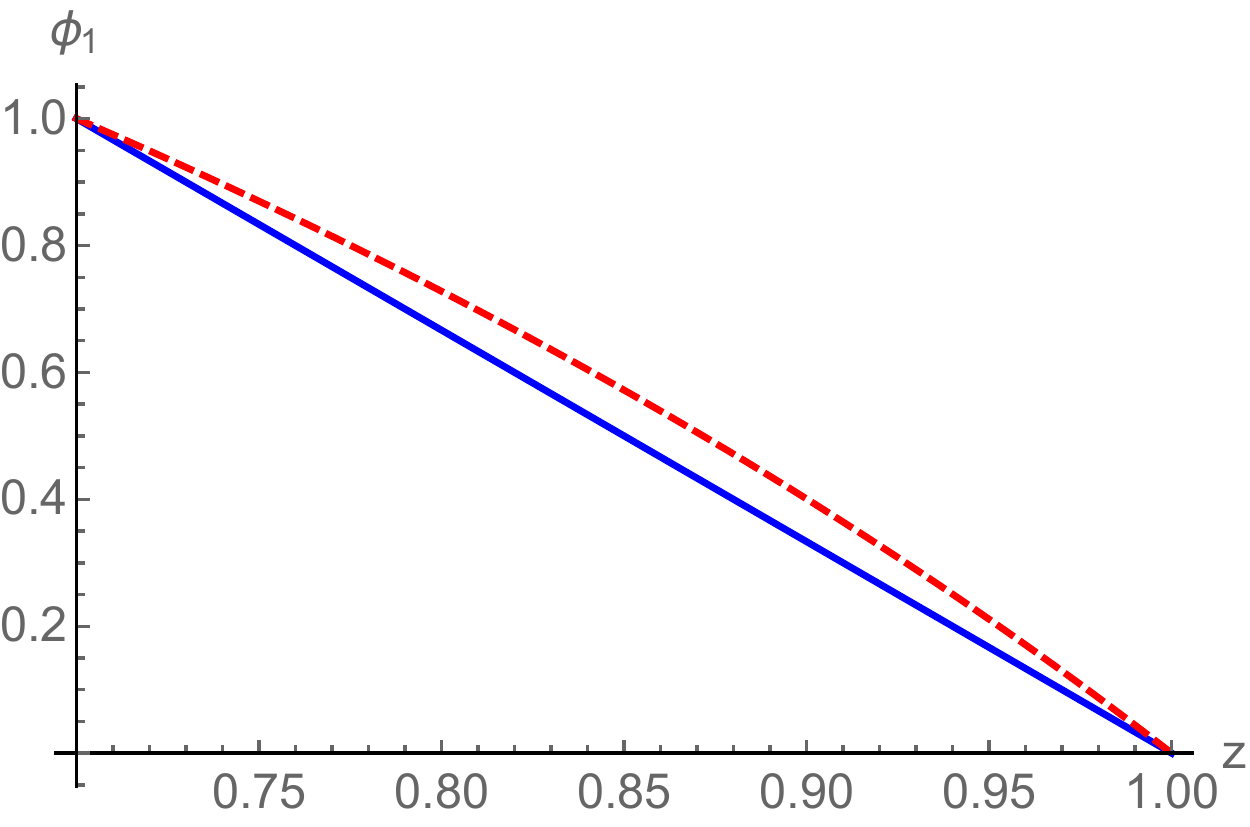}
\includegraphics[height=3.4cm]{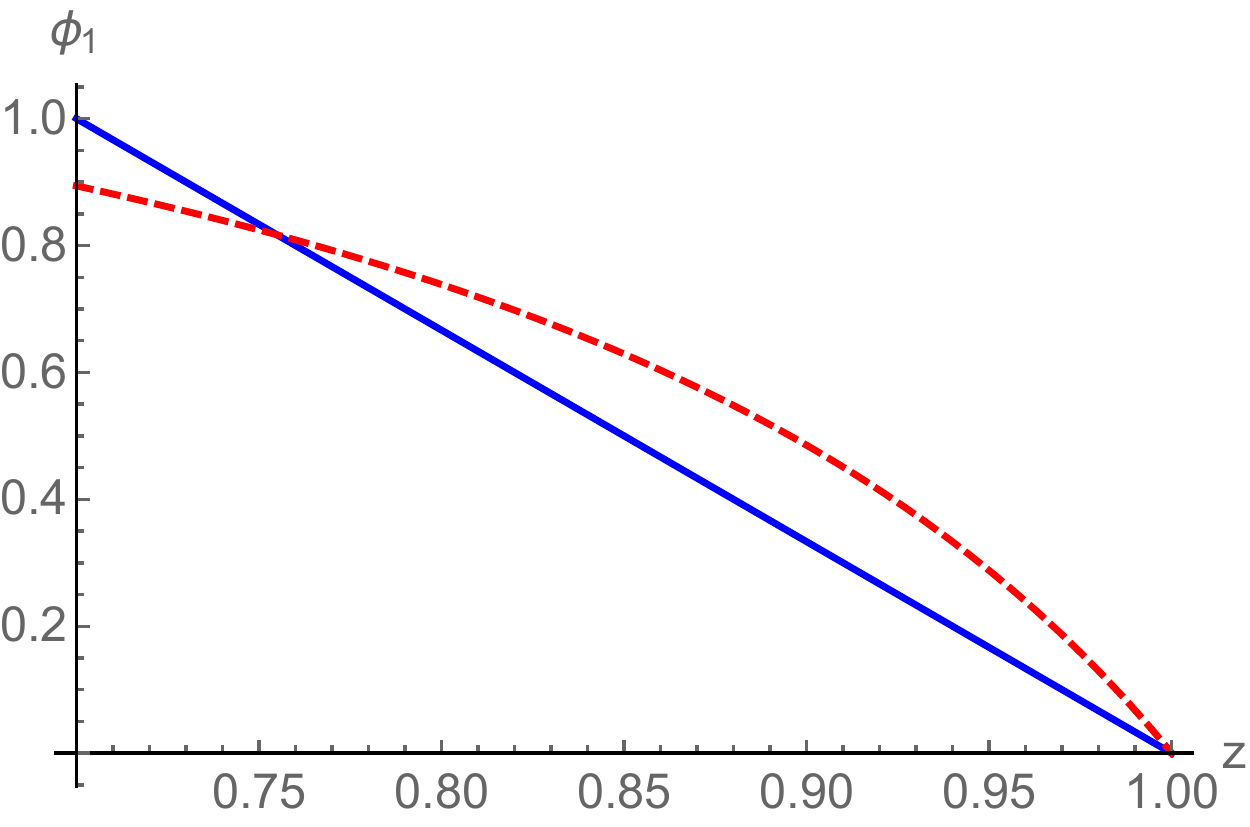}
\caption{Upper-layer modal function $\phi_1$ of interfacial ring waves for $\theta = 0$ (left), $\theta = \frac \pi 2$ (middle) and $\theta = \pi$ (right). Here,  $\rho_1 = 1, \rho_2 = 1.0001, d = 0.7$; $U(z) = \gamma (z-d)$ with $\gamma = 0$ (blue, solid) and $\gamma = 0.015$ (red, dashed). }
\label{fig:phi1}
\end{center}
\end{figure}
 \begin{figure}[h]
\begin{center}
\includegraphics[height=3.4cm]{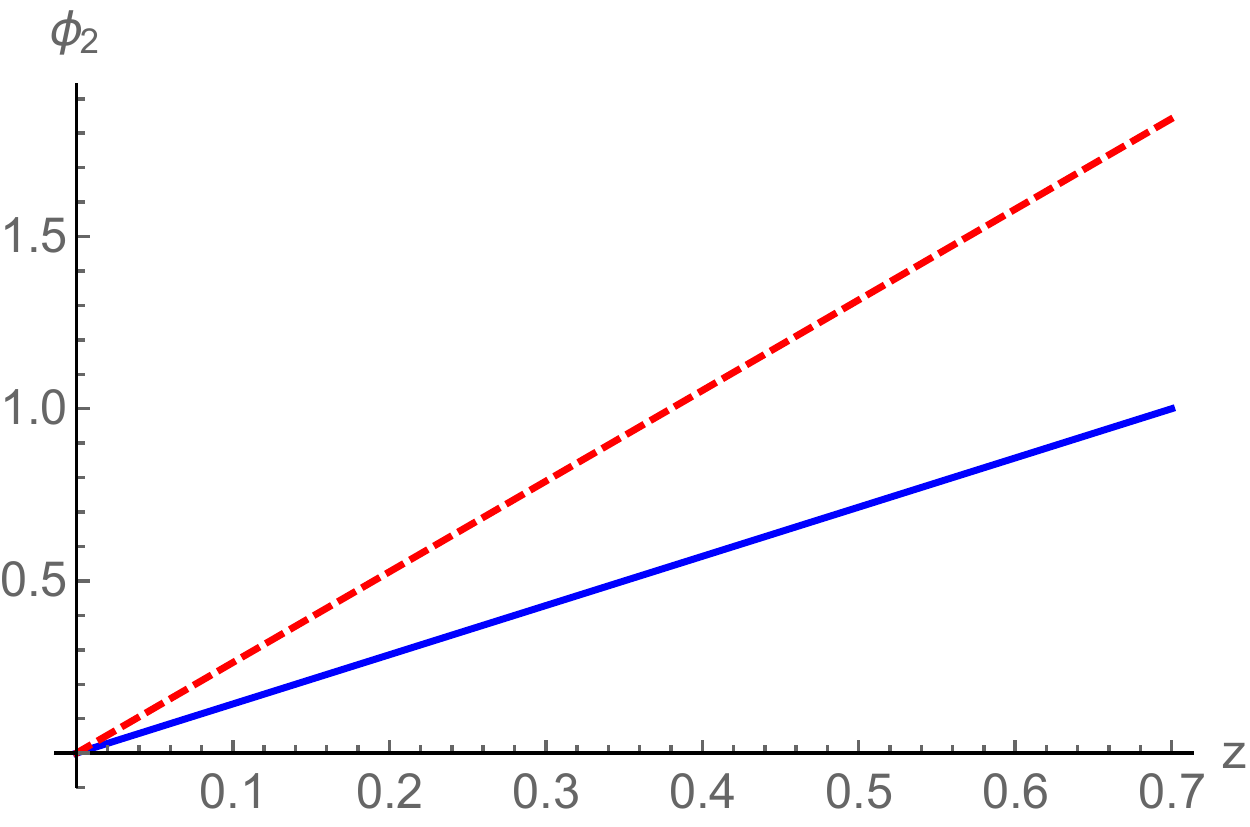}
\includegraphics[height=3.4cm]{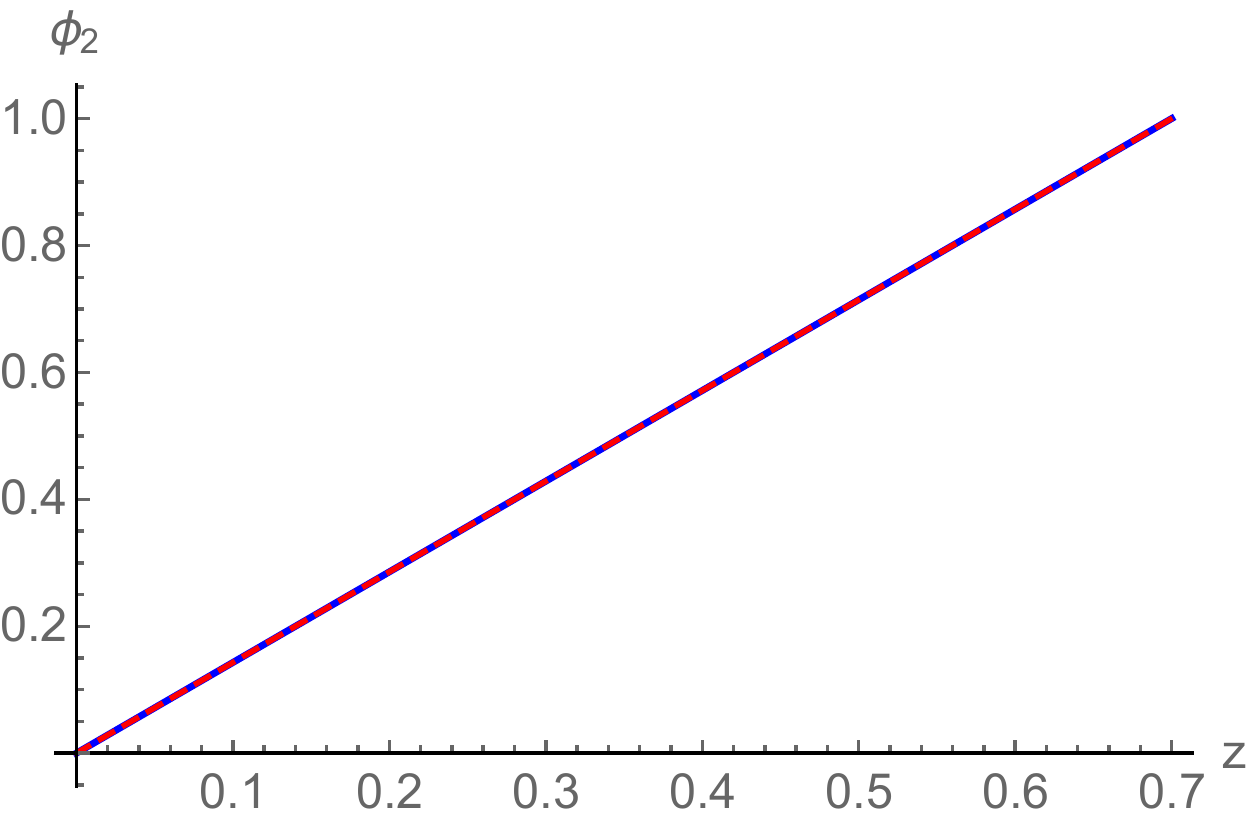}
\includegraphics[height=3.4cm]{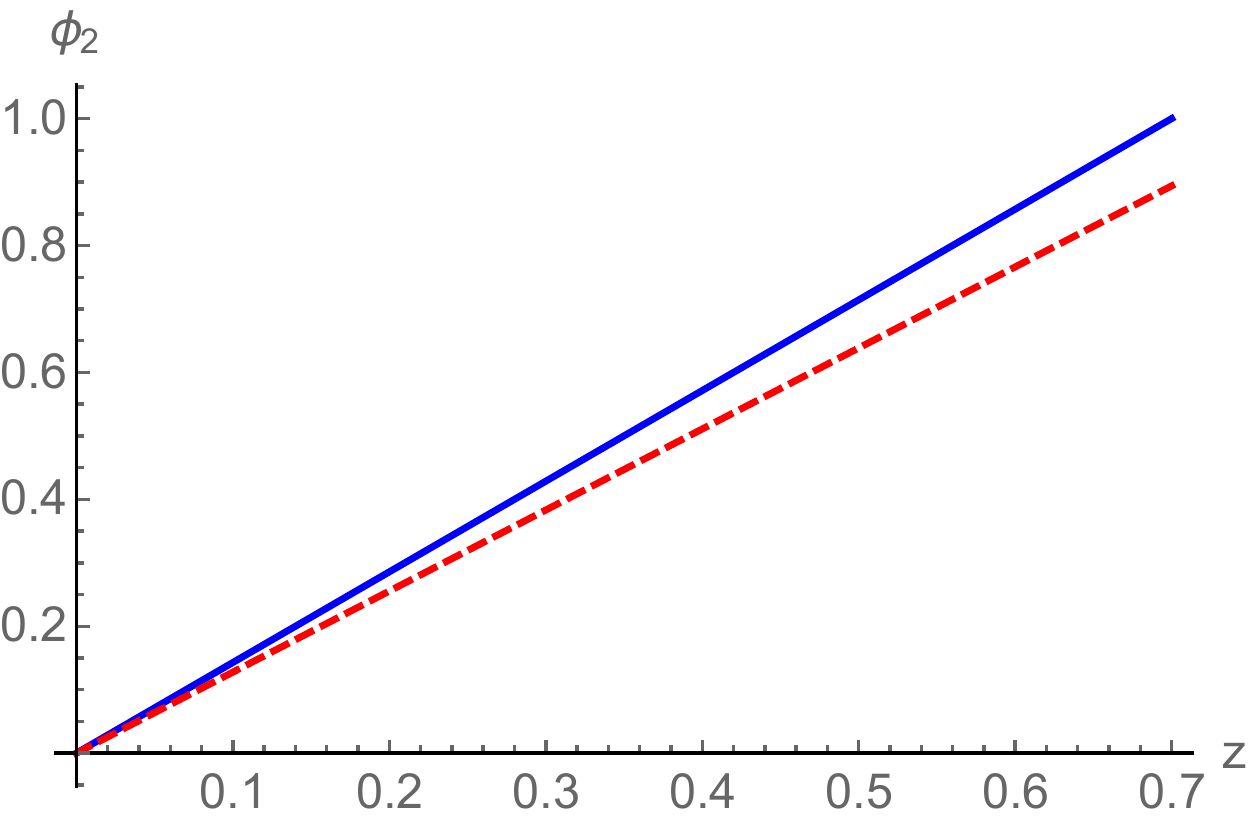}
\caption{Lower-layer modal function $\phi_2$ of interfacial ring waves for $\theta = 0$ (left), $\theta = \frac \pi 2$ (middle) and $\theta = \pi$ (right). Here,  $\rho_1 = 1, \rho_2 = 1.0001, d = 0.7$; $U(z) = \gamma (z-d)$ with $\gamma = 0$ (blue, solid) and $\gamma = 0.015$ (red, dashed). }
\label{fig:phi2}
\end{center}
\end{figure}

It is also instructive to analyse the 3D vertical structure of the internal wave field by illustrating the dependence of the modal functions in the upper and lower layers on $\theta$ and $z$. The plots in Fig.~\ref{fig:phi1}  and Fig.~\ref{fig:phi2}  show the upper- and lower-layer modal functions $\phi_1$ and $\phi_2$, respectively, for the linear current $U(z) = \gamma (z-d)$. The modal functions are given by the formulae (\ref{phi1a}), (\ref{phi2a}), and they have been normalised to be equal to 1 at $\theta = \frac{\pi}{2}$ (orthogonal direction to the current, where the velocity field in the fluid is least affected by the current) and $z = d$ (i.e. on the interface).  Here, $\rho_1 = 1, \rho_2 = 1.0001, d = 0.7$ and  $\gamma = 0$ (blue, solid) or $\gamma = 0.015$ (red, dashed). The upper-layer modal function  $\phi_1(z, \theta)$ is shown in Fig.~\ref{fig:phi1} for three fixed values of $\theta$:  $\theta = 0$ (downstream direction), $\theta = \frac{\pi}{2}$ (orthogonal direction to the current) and $\theta = \pi$ (upstream direction). The lower-layer modal function  $\phi_2(z, \theta)$ is shown in Fig.~\ref{fig:phi2} for the same values of $\theta$.  It is evident that the vertical structure strongly depends on the direction.  For $\gamma = 0.015$, the greatest changes  are in the downstream direction, compared to the case when there is no background shear flow, but there is also considerable variation in the upstream direction. The variation in the direction orthogonal to the current is less significant in the upper layer, and negligible in the lower layer. The upper-layer vertical structure in the downstream direction develops a sharp gradient near the interface.

\begin{figure}[h]
\begin{center}
\includegraphics[height=5cm]{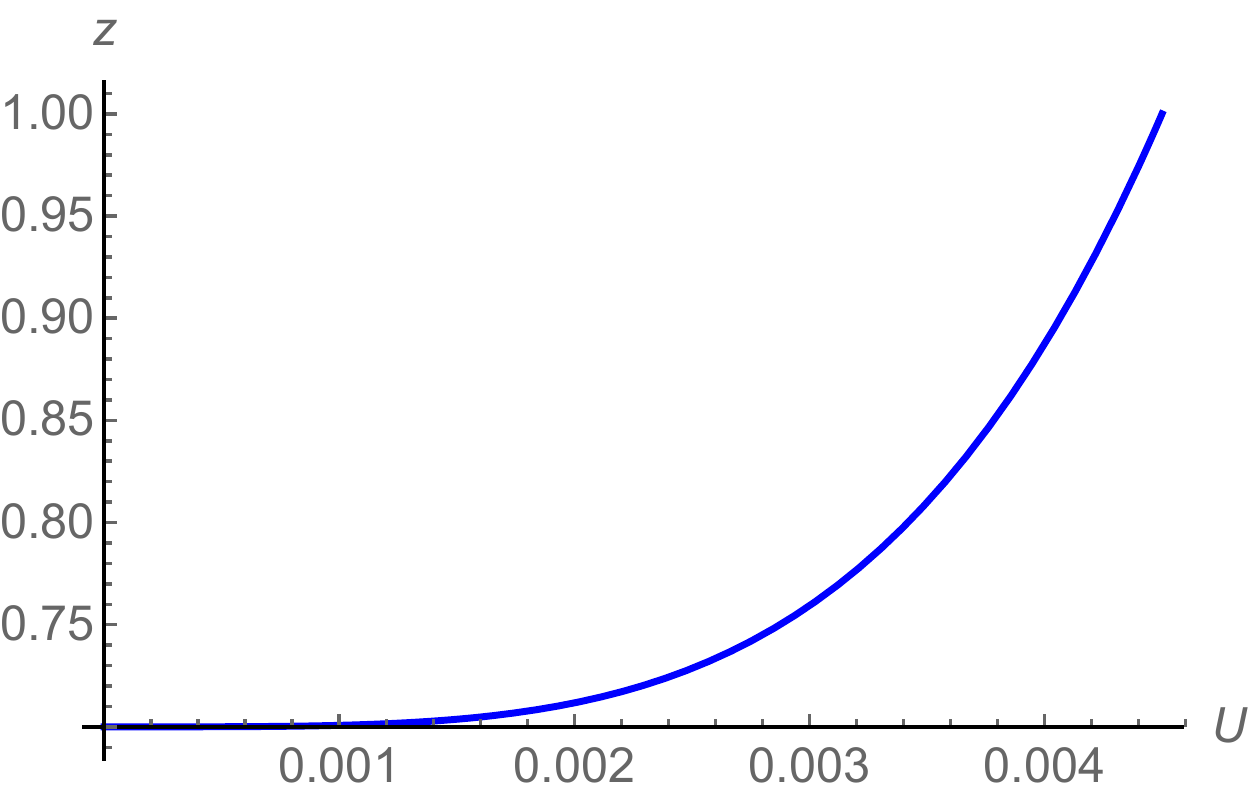}
\includegraphics[height=5.5cm]{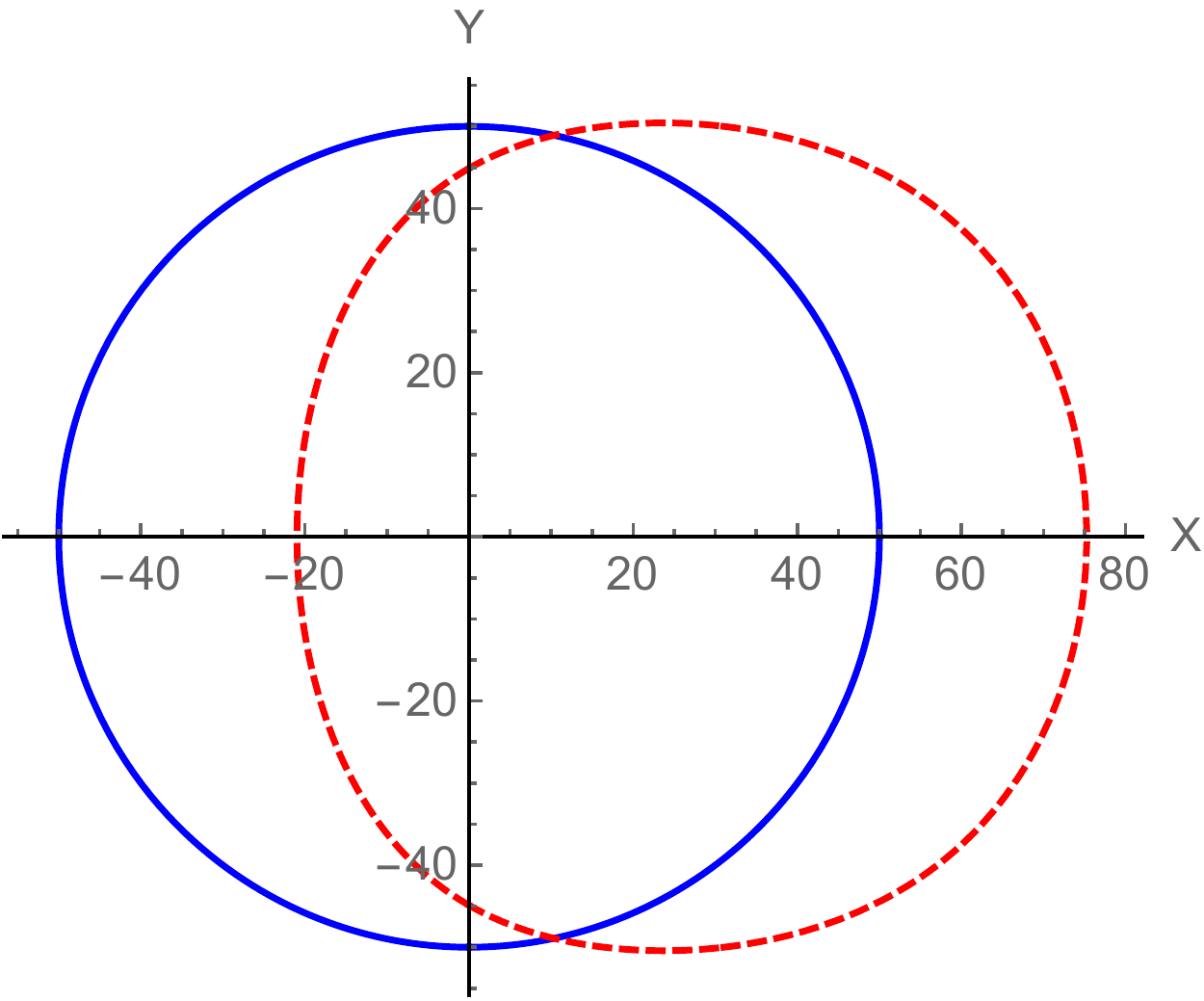}
\caption{Upper-layer current $U(z) = \gamma (z-d)^{1/4}$ for $d=0.7$ and $U(1) =0.0045$ (left), and wavefronts of interfacial ring waves (right) for  $\rho_1 = 1, \rho_2 = 1.0001, d = 0.7$; $\gamma = 0$ (blue, solid) and $\gamma = 0.0060804$ (red, dashed). }
\label{fig:root4}
\end{center}
\end{figure}

 Finally, we note that for the same surface strength $U(1) = 0.0045$ there exist currents in this family which appear to have wavefronts squeezed in the direction of the shear flow, in contrast to the behaviour illustrated in previous plots. Indeed, this happens, for example, for $\alpha = 1/4$ and $\gamma = 0.0060804$ (where we continue to choose $\alpha$ in the form of positive and negative powers of 2). The vertical profile of the current, and the corresponding deformation of the wavefront of the ring wave compared to the case when there is no shear flow are shown in Fig.~\ref{fig:root4}. This behaviour is similar to the squeezing which was previously reported for internal ring waves propagating in a two-layer fluid with the piecewise-constant shear flow \cite{KZ1}, where it was linked to the presence of the long-wave instability for plane waves tangent to the ring at $\theta = 0$ and $\theta = \pi$ for a sufficiently strong current  \cite{O1, O2}.  
 
Indeed,  the upper-layer current $U(z) = \gamma (z-d)^{\alpha}$ tends to $U(z) = \gamma$ as $\alpha \to 0$. In this limit, the equation (\ref{eq_ka}) takes the form
\begin{equation}
k^2 + k'^2 = \frac{\rho_1 d [-s + \gamma (k \cos \theta - k' \sin \theta)]^2 + \rho_2 (1-d) s^2}{(\rho_2 - \rho_1) d (1-d)}
\label{pc}
\end{equation}
and coincides with the equation obtained in \cite{KZ1} for the case of a two-layer fluid with the constant current $U_1$ in the upper layer, and another constant current $U_2$ in the lower layer if we let $U_1 - U_2 = \gamma$. Here, $s^2$ is given by (\ref{sa}), and the general solution has the form (\ref{gs}), where
\begin{equation}
a^2 + b^2 = \tilde \alpha \gamma^2 a^2 - 2 \tilde \alpha \gamma s a + 1, \quad \tilde \alpha= \frac{\rho_1}{(\rho_2-\rho_1) (1-d)} > 0.
\label{conic}
\end{equation}
The singular solution of (\ref{pc}) has been found in \cite{KZ1} under the assumption that the current is sufficiently weak. We now identify three regimes depending on the strength of the current, and provide detailed analysis up to the onset of the long-wave instability. 

If $\gamma^2 < \frac{1}{\tilde \alpha}$ ({\it elliptic regime}, the locus of parameters $a$ and $b$ is a circle)  then the singular solution can be written in the form
\begin{eqnarray}
&&k(\theta)= \sqrt{\frac{1-\tilde \alpha (1 - \tilde \alpha s^2) \gamma^2}{\cos^2 \theta +(1-\tilde \alpha \gamma^2 )\sin^2\theta}}\left (\frac{1 - \tilde \alpha \gamma^2 \sin^2 \theta}{1 - \tilde \alpha \gamma^2} \right ) -\frac{\tilde \alpha  \gamma s}{1-\tilde \alpha \gamma^2} \cos\theta, 
\label{ss1}\\
&& \mbox{where} \quad \theta \in [-\pi, \pi]. \nonumber 
\end{eqnarray}
If $\gamma^2 = \frac{1}{\tilde \alpha}$ ({\it parabolic regime}, the locus of parameters $a$ and $b$ is a parabola), then 
\begin{eqnarray}
&&k(\theta)=  \frac{1}{2 \sqrt{\tilde \alpha} s \cos \theta} (\cos^2 \theta + \tilde \alpha s^2 \sin^2 \theta),
\label{ss2}\\
&& \mbox{where} \quad  \theta \in \left (-\frac{\pi}{2}, \frac{\pi}{2}\right ). \nonumber
\end{eqnarray}
Here, $k(\theta) \to  \infty$ as $\theta \to \pm \frac{\pi}{2}$ leading to the presence of a stationary point on the wavefront at the origin.
Finally, if 
$$\frac{1}{\tilde \alpha} < \gamma^2 < \gamma^2_c = \frac{1}{\tilde \alpha (1- \tilde \alpha s^2)}$$
 ({\it hyperbolic regime}, the locus of parameters $a$ and $b$ is a hyperbola), then the singular solution has two branches (corresponding to the right and left parts of the wavefront):
\begin{eqnarray}
&&k_{r,l}(\theta)= \frac{\tilde \alpha  \gamma s}{\tilde \alpha \gamma^2 -1} \left [ \cos\theta \mp  \frac{\sqrt{1- \tilde \alpha (1- \tilde \alpha s^2) \gamma^2}}{\tilde \alpha  \gamma s} \sqrt{\cos^2 \theta - (\tilde \alpha \gamma^2 - 1 )\sin^2\theta} \right],
\label{ss3}\\
&& \mbox{where} \quad \theta \in  [-\arctan \frac{1}{\sqrt{\tilde \alpha \gamma^2 - 1}},  \arctan \frac{1}{\sqrt{\tilde \alpha \gamma^2 - 1}}].  
\nonumber 
\end{eqnarray}
%Here, $\tilde \alpha= \frac{\rho_1}{(\rho_2-\rho_1) (1-d)} > 0, \quad s^2=\frac{(\rho_2-\rho_1)d (1-d)}{\rho_1 d + \rho_2 (1-d)}$.

It is now instructive to compare the singular solutions (\ref{ss}) and (\ref{ss1}) - (\ref{ss3}),  for $U(z) = \gamma (z-d)$  and  $U(z) = \gamma$, respectively. The solution (\ref{ss}) is real-valued for all values of parameters of the problem, while (\ref{ss1}) - (\ref{ss3}) is real-valued only for 
\begin{equation}
\gamma^2 < 
\gamma^2_{c} = \frac{1}{\tilde \alpha (1- \tilde \alpha s^2)} = \frac{(\rho_2 - \rho_1) [\rho_1 d + \rho_2 (1-d)]}{\rho_1 \rho_2}.
\label{rlcond}
\end{equation}
This condition coincides with the rigid-lid version of the criterion of long-wave instability  for co- and counter-propagating plane waves on that shear flow \cite{O1,O2} (see also \cite{B,BM, LM} and references therein). 

\begin{figure}[h]
\begin{center}
\includegraphics[height=6.5cm]{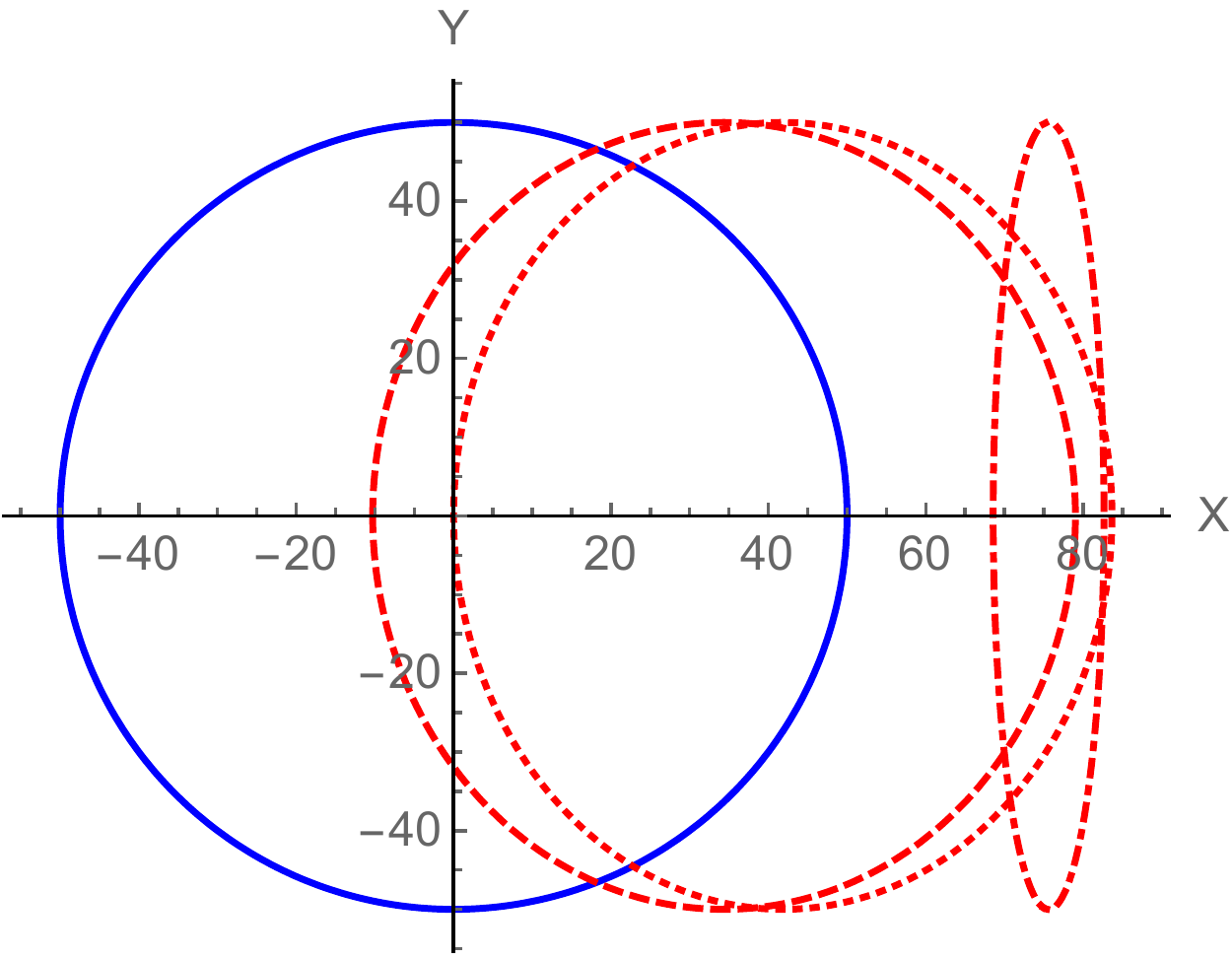}
\caption{Wavefront of an interfacial ring wave for $U(z) = \gamma$, and $\rho_1 = 1, \rho_2 = 1.0001, d = 0.7$; $\gamma = 0.0045 < \frac{1}{\sqrt{\tilde \alpha}} = 0.00547723$ (red, dashed), $\gamma = \frac{1}{\sqrt{\tilde \alpha}}$ (red, dotted), $\frac{1}{\sqrt{\tilde \alpha}}  < \gamma = 0.0099 < \gamma_c = 0.00999965$ (red, dot-dashed). }
\label{fig:squeezing}
\end{center}
\end{figure}

The deformation of the wavefront of a ring wave for the increasing strength of the current is shown in Fig.~\ref{fig:squeezing}. As the strength of the current $\gamma$ approaches the critical value $\gamma_c$, the wavefront appears to be more and more squeezed, until the ring collapses at the critical value. In the elliptic regime a part of the wavefront propagates in the upstream direction, in the parabolic regime the wavefront has a stationary point at the origin, while in the hyperbolic regime all parts of the wavefront propagate in the downstream direction, and for each value of $\theta$ in the interval $\theta \in [-\arctan \frac{1}{\sqrt{\tilde \alpha \gamma^2 - 1}}, \arctan \frac{1}{\sqrt{\tilde \alpha \gamma^2 - 1}}]$  the singular solution has two branches.

Since the family of currents approaches the piecewise-constant current as $\alpha \to 0$, we anticipate that the squeezing of the wavefront of the ring wave will persist for the values of $\alpha$ smaller than a certain threshold value. This observation invites studies of the long-wave instability of the plane and ring waves propagating on the background of such currents, but this is beyond the scope of our present paper.

\section{Conclusion}

In this paper we considered long internal ring waves in a two-layered fluid with a rather general depth-dependent parallel upper-layer current in the rigid-lid approximation. Our main aim was to obtain an analytical solution of the nonlinear first-order ordinary differential equation responsible for the adjustment of the speed of the long interfacial ring waves in different directions, which was key to the subsequent analysis of the wave field. For a large family of the upper-layer current profiles described by the function
$$
U(z) = \gamma (z-d)^{\alpha},
$$
where $d$ is the position of the interface, while $\gamma$ and $\alpha$ are some positive constant parameters, an explicit analytical solution was obtained in terms of the hypergeometric function. In many natural cases the solution reduces to elementary functions, and we considered examples with $\alpha =  2, 1, \frac 12$ and $ \frac 14$ (in the latter case the corresponding formula in terms of elementary functions is rather long, therefore we did not show it). 

The constructed solution has allowed us to illustrate the effects of such shear flows on the wavefronts and vertical structure of interfacial ring waves with an emphasis on the effects of the density jump and the type and the strength of the current. For the same strength of the shear flow, all flows had greater effect on internal waves for smaller values of the density jump (i.e. slower internal waves). For the currents with $\alpha = 1, 2, \frac 12$ with the same strength on the surface, the current with $\alpha = \frac 12$ convected the ring further downstream than both other currents, while the current with $\alpha = 2$ had greater effect on the shape of the wavefront than both other currents, with all currents elongating the ring in the direction of the current. The $\alpha=\frac 12$ current and other currents with $\alpha < 1$ could be more representative of the river inflows and exchange flows in straits, while the $\alpha = 2$ and other currents with $\alpha > 1$ could be closer to the wind-generated currents. While all currents had the same strength on the surface, the variation of their behaviour in the bulk of the layer had a profoundly different effect on the propagation of the ring waves. The vertical structure of the internal wave field is strongly three-dimensional with the greatest changes due to the current in both layers being in the downstream direction. 

We also showed that for the same surface strength, the current with $\alpha = \frac 14$ leads to the squeezing of the wavefront of the ring wave in the direction of the current, similarly to the previously reported behaviour for the ring waves on a piecewise-constant current, which we now revisited and described in detail up to the onset of the long-wave instability.  We conjecture that this behaviour in the stable regime is generally indicative of the presence of the long-wave instability for sufficiently strong currents in our family with $\alpha$ less than some threshold value (possibly, $\alpha < \frac 12$)  which we  hope to address in our future studies. 

\section{Acknowledgments}

I would like to use this opportunity to thank Efim Pelinovsky for years of support, advice and friendship.

%\newpage

\end{document}